\documentclass[aps,prb,twocolumn,superscriptaddress,10pt]{revtex4-1}

\usepackage{graphicx}
\usepackage{amssymb}
\usepackage{amsmath}
\usepackage{color}
\usepackage{hyperref}

\begin{document}
\newcommand*{\beq}{\begin{equation}}
\newcommand*{\eeq}{\end{equation}}
\newcommand*{\balign}{\begin{align}}
\newcommand*{\eall}{\end{align}}
\newcommand*{\bitem}{\begin{itemize}}
\newcommand*{\eitem}{\end{itemize}}
\newcommand*{\benum}{\begin{enumerate}}
\newcommand*{\eenum}{\end{enumerate}}
\newcommand*{\vecr}{\mathbf{r}}
\newcommand*{\vecR}{\mathbf{R}}
\newcommand*{\veck}{\mathbf{k}}
\newcommand*{\vecq}{\mathbf{q}}
\newcommand*{\vecS}{\mathbf{S}}
\newcommand*{\vecx}{\mathbf{x}}
\newcommand*{\unitvece}{\hat{\mathbf{e}}}
\newcommand*{\DTO}{Dy$_2$Ti$_2$O$_7$}
\newcommand*{\HTO}{Ho$_2$Ti$_2$O$_7$}
\newcommand*{\DMSO}{Dy$_3$Mg$_2$Sb$_3$O$_{14}$}

\title{Supercooling and fragile glassiness in a dipolar kagome Ising magnet}

\author{James Hamp}
\email[]{james.o.hamp@gmail.com}
\affiliation{TCM Group, Cavendish Laboratory, University of Cambridge, J.~J.~Thomson Avenue, Cambridge CB3 0HE, United Kingdom}

\author{Roderich Moessner}
\affiliation{Max-Planck-Institut f\"ur Physik komplexer Systeme, N\"othnitzer Stra{\ss}e 38, Dresden 01187, Germany}

\author{Claudio Castelnovo}
\affiliation{TCM Group, Cavendish Laboratory, University of Cambridge, J.~J.~Thomson Avenue, Cambridge CB3 0HE, United Kingdom}

\date{\today}

\begin{abstract}
We study equilibration and ordering in the classical dipolar kagome Ising antiferromagnet, which we show behaves as a disorder-free fragile spin glass. 
By identifying an appropriate order parameter, we demonstrate a transition to the ordered state proposed by Chioar \emph{et al.} [Phys.~Rev.~B \textbf{93}, 214410 (2016)] with a 12-site unit cell that breaks time-reversal and sublattice symmetries, and further provide evidence that the nature of the transition is first order.
Upon approaching the transition, the spin dynamics slow dramatically. The system readily falls out of equilibrium, overshooting the transition and entering a supercooled liquid regime. 
Using extensive Monte Carlo simulations, we show that the system exhibits super-Arrhenius behavior above the ordering transition. The best fit to the relaxation time is of the Vogel-Fulcher form with a divergence at a finite ``glass transition'' temperature in the supercooled regime. Such behavior, characteristic of fragile glasses, is particularly remarkable as the model is free of quenched disorder, does not straightforwardly conform to the avoided criticality paradigm, and is simple and eminently realizable in engineered nanomagnetic arrays.  
\end{abstract}

\maketitle
%
%

\section{Introduction}

It is well known that the presence of disorder in a system can generate a rugged free-energy landscape resulting in slow or glassy dynamics. 
The converse---the appearance of glassiness in the absence of disorder---is far less understood. 
Due to their complex energy landscapes exhibiting multiple minima, geometrically frustrated systems with long-range interactions are natural candidates in this regard~\cite{Tarjus2005,Dzero2012}. 
In recent years, interesting glassy slow dynamics in the absence of disorder has been uncovered in electronic Coulomb liquids on the triangular lattice at quarter-filling~\cite{mahmoudian_glassy_2015}. 
The dynamics of electrons on the frustrated kagome lattice has also received some attention of late~\cite{terao_hopping_2016}, but strong metastability effects mean there remain open questions about the nature of the ground state in that system. 
Slow dynamics persist even for faster-decaying interactions (dipolar instead of Coulomb) in systems without particle-number conservation, i.e., spin systems. This was demonstrated, for instance, in Ref.~\onlinecite{chioar_ground-state_2016} where due to strong freezing and metastability effects the nature of the ground state could not convincingly be established. 

In this paper, we explore in greater detail the latter case, namely, that of frustrated Ising spins on the kagome lattice subject to dipolar interactions---the dipolar kagome Ising antiferromagnet (DKIAFM). 
We begin by identifying an order parameter for the state proposed by Chioar \emph{et al.}~\cite{chioar_ground-state_2016}. 
This allows us to confirm the nature of the ground state in extensive simulations of small systems and to provide evidence that the nature of the transition is first order.
Approaching the transition, the spin dynamics slow dramatically, and a supercooled liquid regime can appear upon further cooling. Despite the propensity of the system to fall out of equilibrium, we are able to reach thermodynamic equilibrium in Monte Carlo simulations for systems of up to around $300$~spins. 
At equilibrium above the ordering transition we find robust evidence of super-Arrhenius behavior. The relaxation time $\tau$ appears to diverge according to a Vogel-Fulcher law $\tau \sim \exp[\Delta/(T-T_0)]$, characteristic of fragile glasses, at a temperature lower than the thermodynamic transition temperature. 
The glassy slowing down may be related to the existence of many low-lying metastable states exhibiting dendritic arrangements of emergent charges. 


Our results highlight the DKIAFM as particularly suitable for the study of disorder-free glassy dynamics. On one hand, fragile glass behavior where the timescale diverges at a finite temperature in the supercooled liquid regime, is hard to come by in nondisordered systems in two dimensions~\footnote{Fragile glass behavior in higher-dimensional systems in the presence of long-range interactions has been uncovered in Refs.~\onlinecite{Grousson2002,Grousson2002a}, for example. However, the computational challenge of accessing large timescales prevented the authors from being able to analyse the super-Arrhenius temperature regime---a task that is just within reach in two-dimensional systems.}. 
On the other hand, theoretical models of glasses without disorder where the thermodynamic behavior is well understood are typically limited to the rather artificial multispin Hamiltonians of kinetically constrained models~\cite{Ritort2003}, difficult to realize in a laboratory, and unlikely to occur in real materials. The DKIAFM exhibits the above interesting features with a Hamiltonian that is eminently realistic. 

Experimental realizations of the DKIAFM have already been obtained 
using artificial nanomagnetic arrays~\cite{Chioar2014} (indeed it is these realizations that motivated the authors of Ref.~\onlinecite{chioar_ground-state_2016} to first study this model), and one may be able to study similar systems in real time on a ``microscopic'' scale~\cite{nisoli_textitcolloquium_2013}. 
The behavior exhibited by the DKIAFM may also be relevant to monolayer colloidal crystals~\cite{han_geometric_2008,zhou_glassy_2017} 
where recent advances have enabled the study of slow dynamics of frustrated systems in real time. 
Another potential avenue where this model could be realized in experiments is that of cold polar molecules~\cite{Ni2008} and atomic gases with large magnetic dipole moments~\cite{Griesmaier2005}. Of particular interest there would be the possibility of investigating how the relaxation time scales and glassy behavior may be affected by quantum dynamics. 
Finally, (layered) kagome spin lattices occur in solid-state materials and can be combined with the crystal-field physics of rare-earth ions to achieve the desired easy-axis (Ising) nature and leading dipolar interactions~\cite{Scheie2016,Paddison2016,Dun2017}. 

The remainder of the paper is organized as follows. 
In Sec.~\ref{sec:model} we introduce the DKIAFM model. 
In Sec.~\ref{sec:ground_state} we discuss the ordering displayed by the model and the nature of the transition to the ground state.
In Sec.~\ref{sec:spin_relaxation} we discuss the dynamic properties of the model at low temperatures, in particular, the spin relaxation. 
In Sec.~\ref{sec:out_of_eq} we consider the dynamics of the model out of equilibrium.
Finally in Sec.~\ref{sec:conclusion} we conclude and discuss the connection to experiments.
%
%

\section{Model}
\label{sec:model}

\begin{figure}[t!]
\centering	\includegraphics[width=1.0\linewidth]{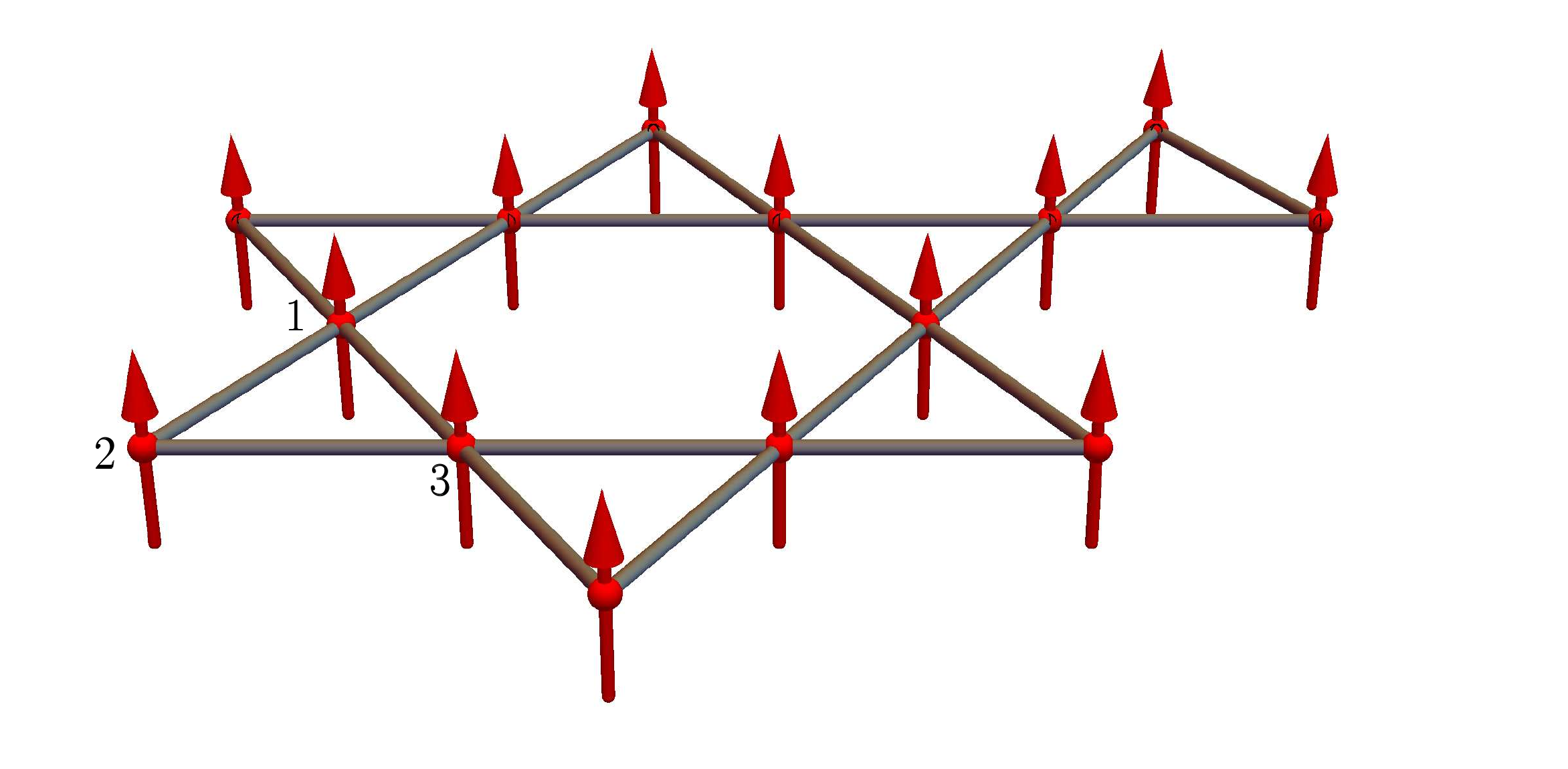}
\caption{\label{fig:kagome_lattice_111} 
	The dipolar kagome Ising antiferromagnet.
	The spins $\mathbf{S}_i$ (red arrows) share a global Ising easy axis perpendicular to the plane (in the $\hat{\mathbf{e}}_z$ direction) and interact via nearest-neighbor exchange interactions of strength $J$ and long-range dipolar interactions of characteristic strength $D$. 
	The sublattices $a = \{ 1, 2, 3\}$ are numbered.
}
\end{figure}

We consider the dipolar kagome Ising antiferromagnet~\cite{chioar_ground-state_2016} (illustrated in Fig.~\ref{fig:kagome_lattice_111}) in which $N$ classical spins $\mathbf{S}_i$ are arranged on the kagome lattice and share a global Ising easy axis perpendicular to the plane (in the $\hat{\mathbf{e}}_z$ direction). 
As such, the spins $\mathbf{S}_i= \mu \sigma_i \hat{\mathbf{e}}_z$ can be described by the Ising pseudospin variables $\{ \sigma_i = \pm 1 \}$, and $\mu$ is the magnitude of the spin magnetic moment (we set $\mu = 1$ where relevant in the remainder of the paper). 

The Hamiltonian comprises an exchange term of strength $J$ between spins at nearest-neighbor lattice sites $\langle i j \rangle$ and long-range dipolar interactions of characteristic strength $D=(\mu_0/4 \pi) \mu^2 /r_{\text{nn}}^3$ between all pairs of spins, where $r_{\text{nn}}$ is the nearest-neighbor distance of the lattice.
The Hamiltonian is given by
\begin{equation}
	\mathcal{H}  =  
	-J \sum_{\langle ij \rangle} \sigma_i \sigma_j 
	+D r_\text{nn}^3 \sum_{j>i} \frac{\sigma_i\sigma_j}{|\mathbf{r}_{ij}|^3},
\label{eq:dkiafm_hamiltonian}
\end{equation}
where $\mathbf{r}_{ij} \equiv \mathbf{r}_j - \mathbf{r}_i$ is the separation between spins at lattice sites $i$ and $j$.
The Hamiltonian~\eqref{eq:dkiafm_hamiltonian} is equivalent to dipolar spin ice on the kagome lattice (see, e.g.,~Ref.~\onlinecite{chern_two-stage_2011}) in the limit of spins rotated so that they are perpendicular to the plane. 

We are interested here in the case where both interactions are antiferromagnetic, namely, $J<0$ and $D>0$. The case where $D=0$ is known to be fully frustrated and does not order down to zero temperature~\cite{Takagi1993}. The phase diagram of the case where $J=0$ is less well understood but the system is again strongly frustrated with any ordering (if present at all) suppressed down to temperatures $T \ll D$~\cite{Chioar2014}. A more detailed discussion of the frustration in Eq.~\eqref{eq:dkiafm_hamiltonian} is given in Appendix~\ref{app:quantifying_frustration}.
Throughout the remainder of this paper we consider the coupling parameters from Ref.~\onlinecite{chioar_ground-state_2016}, namely, \mbox{$D=1$}~K and \mbox{$J=-0.5$}~K (we set $k_\text{B} =1$ and measure all energies in Kelvin).
%
%

\section{Order parameter and nature of the transition} 
\label{sec:ground_state}

\begin{figure}[t!]
\centering	\includegraphics[width=1.0\linewidth]{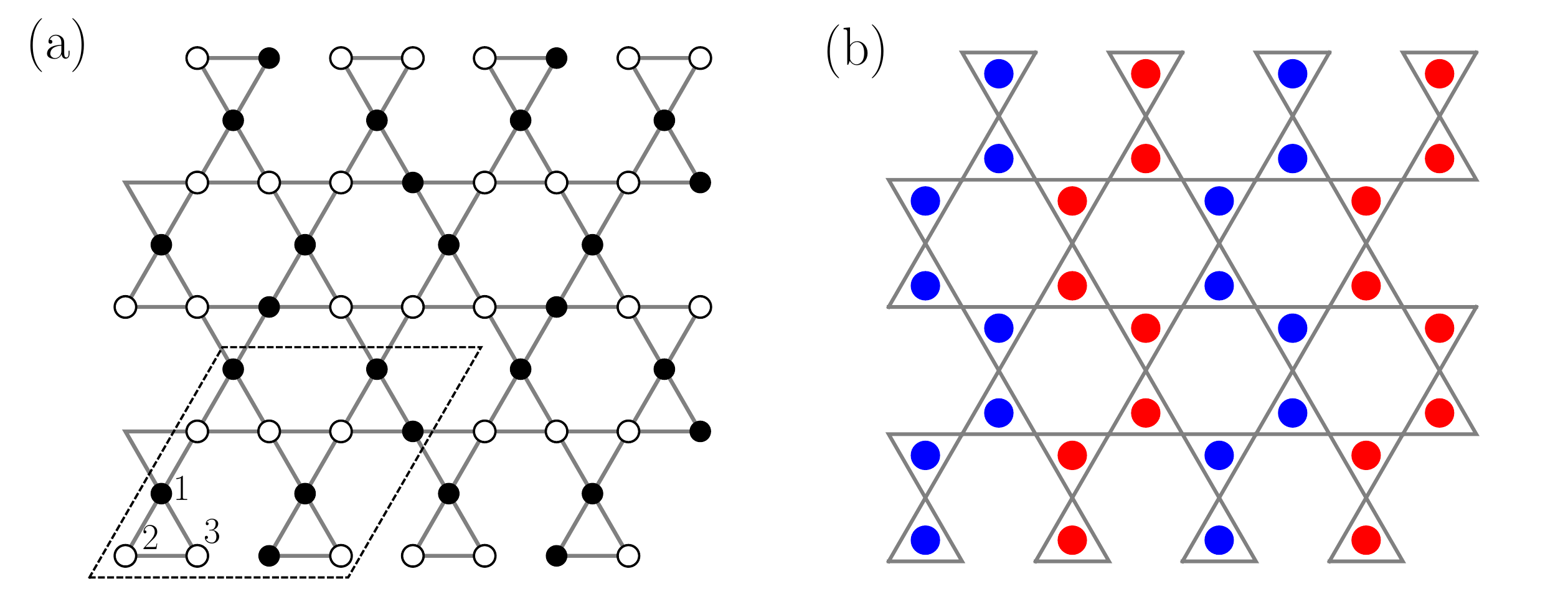}
\caption{\label{fig:7_shape_state}  
Proposed ground state of the dipolar kagome Ising antiferromagnet. 
(a) The spins exhibit a pattern which breaks both time-reversal and sublattice symmetries with one sublattice polarized (here, sublattice 1). 
The 12-site magnetic unit cell is outlined (rhomboid box). 
Black points indicate up spins ($\sigma = 1$), and white points indicate down spins ($\sigma=-1$). 
(b) The charges $Q$ (defined in the text) exhibit a charge-stripe pattern: red dots indicate positive charges ($Q=1$), and blue dots indicate negative charges ($Q=-1$). 
	}
\end{figure}

Long-range ordering has not yet been directly observed in the DKIAFM.
In Ref.~\onlinecite{chioar_ground-state_2016} it was found that at very low temperatures the system exhibits freezing of single spin flip and loop dynamics while seemingly being on the verge of an ordering transition as evidenced by the onset of a pronounced peak in the specific heat. 
A candidate for the ground state of the present model was proposed and shown to be consistent with the available thermodynamic data from simulations.
The state, illustrated in Fig.~\ref{fig:7_shape_state}, has a 12-site magnetic unit cell that can be visualized as arising from tesselating trapezoids of alternating spins to form {\sf 7} shapes. 
For more details on this construction see Ref.~\onlinecite{chioar_ground-state_2016}; in the following we refer to this state as the proposed ground state. 

The state is sixfold degenerate---under threefold rotation (sublattice) symmetry and twofold time-reversal symmetry. Upon assigning an emergent charge variable to each of the up- and down-type triangles ($\bigtriangleup$ and $\bigtriangledown$) on the kagome lattice,
\begin{align}
Q_{\bigtriangleup}=\sum_{i \in \bigtriangleup} \sigma_i; \\ 
Q_{\bigtriangledown}=-\sum_{i \in \bigtriangledown} \sigma_i,
\end{align}
the proposed ground state can be seen to exhibit a charge-stripe pattern [see Fig.~\ref{fig:7_shape_state}(b)].
Here we confirm the proposed low-temperature ordered state and study in detail the thermodynamic behavior of the system by devising an appropriate order parameter and performing extensive Monte Carlo simulations that manage to achieve thermodynamic equilibrium below the ordering temperature. 

Upon inspecting the  spin configuration in Fig.~\ref{fig:7_shape_state}(a), we note that one sublattice of the kagome triangles (in this instance, sublattice 1) is completely polarized with the state having zero magnetization overall. 
This observation leads us to postulate that the proposed ground state, which breaks time-reversal and sublattice symmetries, can be described by an appropriate order parameter for the transition, namely, the sublattice magnetization,
\begin{equation}
m_a = \frac{3}{N} \sum_{i \in \text{sublat.}\, a} \sigma_i, 
\label{eq:sublattice_magnetisation}
\end{equation}
where $a \in \{1,2,3\}$.
The proposed ground state has one sublattice $a \equiv a'$ polarized such that \mbox{$m_{a'}=\pm 1$} 
and the other two sublattices with \mbox{$\sum_{a \neq a'} m_{a} = \mp 1$}, such that the state has zero magnetization overall. 
Notice the spin pattern on the two non-polarized sublattices: it has period four with three spins $\sigma = \mp 1$ followed by one spin $\sigma = \pm 1$ [along the horizontal bonds in Fig.~\ref{fig:7_shape_state}(a)]; such lines of spins are stacked in a specific chiral structure. 

To verify this, we performed extensive Monte Carlo (MC) simulations of the DKIAFM Hamiltonian~\eqref{eq:dkiafm_hamiltonian} using a 12-site unit cell commensurate with the proposed ground state. 
Our system  contains  $N=12L^2$ spins. 
To ensure that we do not exclude other possible ordered states, we also considered system sizes that are commensurate with plausible competing phases, which include the $\sqrt{3} \times \sqrt{3}$ state~\footnote{The unit cell chosen for our system is not commensurate with all six proposed {\sf 7} shape ground states, given by the twofold time-reversal and threefold sublattice symmetries. Indeed, it is commensurate with only one of the three sublattices being fully polarized.}.
Unlike Chioar \emph{et al.}~(see Ref.~\onlinecite{chioar_ground-state_2016}) and Chioar (see Ref.~\onlinecite{chioar_thesis}), we sum the pairwise dipolar interactions via the method of summation of copies employed in Refs.~\onlinecite{moller_magnetic_2009,chern_two-stage_2011} until convergence of one part in $10^{6}$. 
Since loop dynamics do not appear to help in alleviating the freezing~\cite{chioar_ground-state_2016}, we use Metropolis single spin-flip dynamics throughout. 
We cool the system from equilibrium at \mbox{$T/D=1$} in increments of \mbox{$0.5 \times 10^{-3} \, D$} using $2 \times 10^4$ modified MC sweeps for equilibration at each temperature step. Following Ref.~\onlinecite{chioar_ground-state_2016}, we take a modified Monte Carlo sweep at a given temperature to correspond to $N \times r^{-1}$ single spin-flip attempts, where $r$ is the acceptance ratio at that temperature~\footnote{At the very lowest temperatures, the acceptance rate becomes extremely small ($r<10^{-4}$), and it becomes unfeasible to adhere to this protocol. We therefore impose a cutoff $r_\textrm{cutoff}=\textrm{max}[r,10^{-4}]$. We find that such a cutoff does not affect the ability of sufficiently small systems to order, and, in fact, it kicks in below the transition for the system sizes we treat here.}. 
	We time average at each temperature by measuring quantities 400 times, each measurement separated by 50 modified MC sweeps, and we ensemble average the results over 64 independent simulations. 
We note that this simulation protocol requires a substantial investment of computational resources but, by careful analysis of spin autocorrelation functions, we are able to ensure thermodynamic equilibrium down to temperatures lower than the transition temperature, at least for sufficiently small system sizes as discussed below. 

\begin{figure}[t!]
\centering
\includegraphics[width=1.0\linewidth]{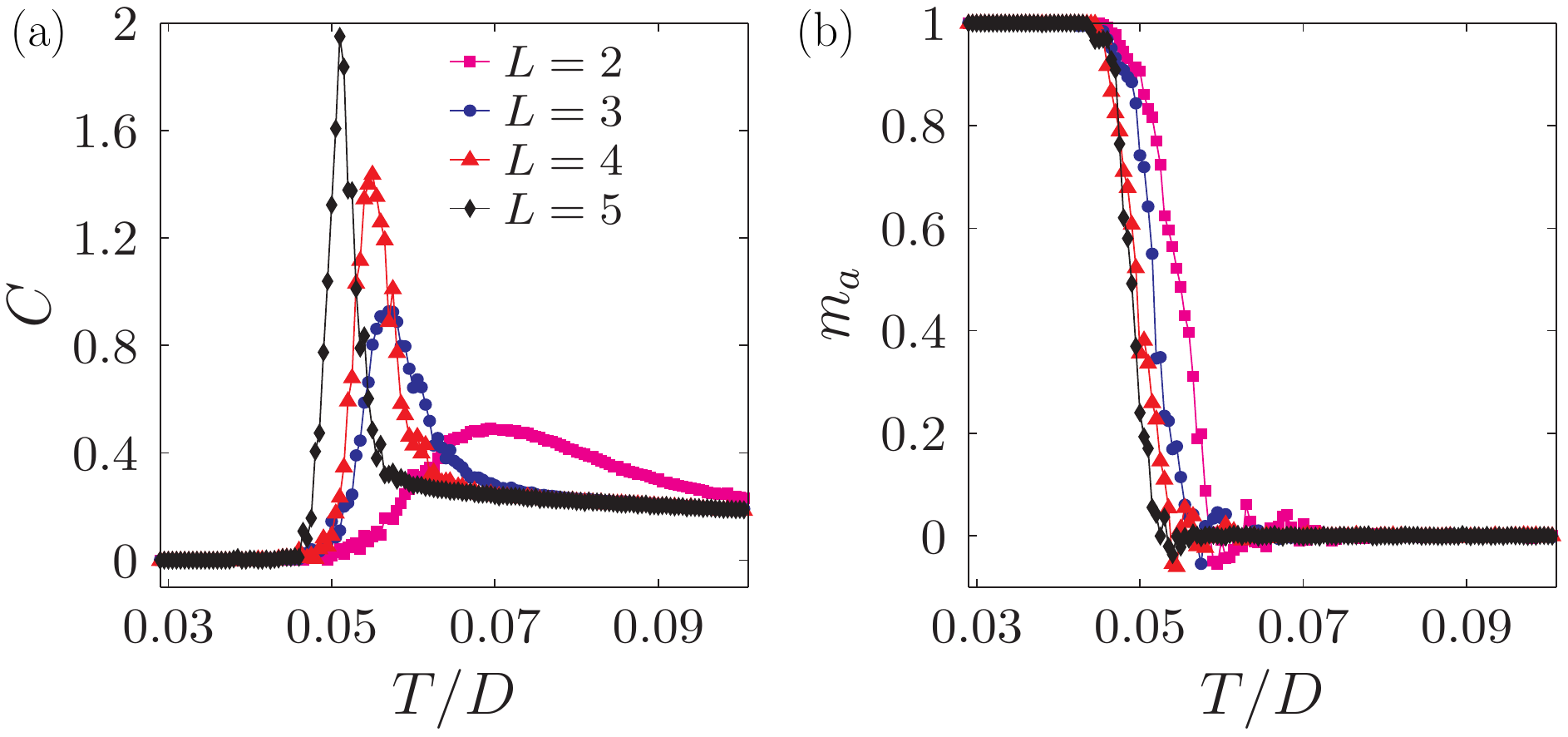}
\caption{\label{fig:thermo_111_inset} 
(a) Specific heat $C$ and (b) sublattice magnetization $m_a$ of the DKIAFM for system sizes \mbox{$L=\{2,3,4,5\}$}. 
The ordering transition to the proposed ground state is signaled by a peak in the specific heat $C$ and concomitant increase in the sublattice magnetization $m_a$ which acts as an order parameter. 
	}
\end{figure}

The specific heat $C$ and sublattice magnetization $m_a$ are shown in Figs.~\ref{fig:thermo_111_inset}(a) and \ref{fig:thermo_111_inset}(b), respectively, for system sizes $L=\{2,3,4,5\}$ with $N=\{48,108,192,300\}$ spins.
The ordering transition is signaled by a peak in the specific heat (at around \mbox{$T_\text{c}/D \simeq 0.05$} for $L=5$) and a clear concomitant increase in $m_a$ from zero to one, signaling the complete development of order consistent with the proposed ground state.
Direct inspection of the spin configurations confirms that indeed the system in each case reaches the proposed ground state.
The freezing of spin dynamics at low temperatures is remarkably strong, and we were unable to fully equilibrate systems larger than $L=5$ (300 spins) in times compatible with also obtaining enough data for averaging purposes. 

\begin{figure}[tp]
\centering	\includegraphics[width=1.0\linewidth]{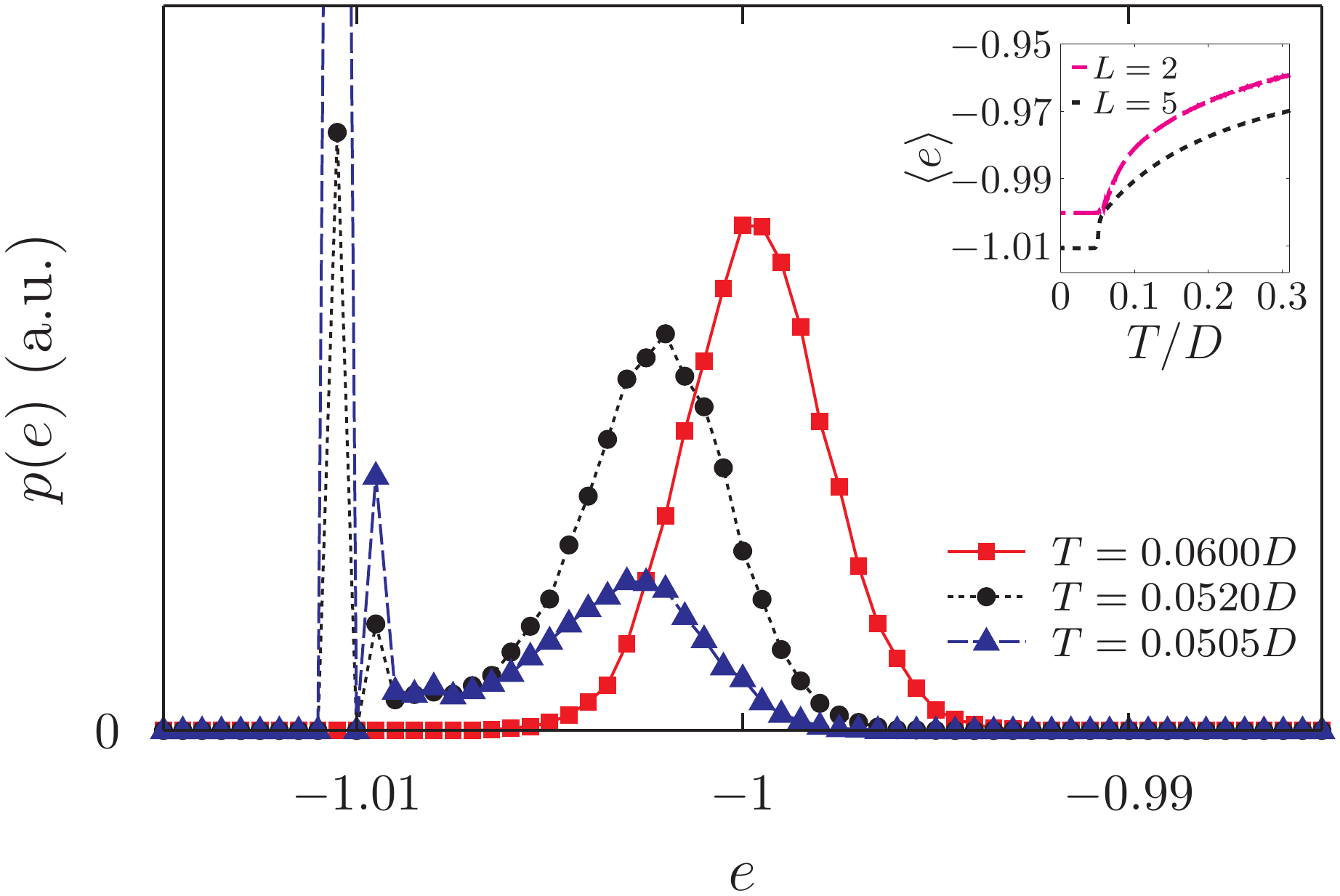}
\caption{\label{fig:energy_111_inset}
Energy histogram $p(e)$ around the transition temperature for $L=5$ (the largest system size we are able to fully equilibrate). 
Just above the transition temperature ($T/D=0.060$), a single Gaussian-like peak indicates the presence of a unique phase. 
Around the transition temperature ($T/D=0.052$), $p(e)$ displays a double-peaked structure indicating the coexistence of two distinct phases (one of which is the ground state) and thus a first-order transition. 
At lower temperatures ($T/D=0.0505$), the higher-energy peak becomes comparatively much less pronounced as the system increasingly occupies the low-energy state. 
Inset: average energy per spin $\langle e \rangle$ as a function of temperature $T$ for two different system sizes $L= \{ 2,5 \}$. 
The transition is signaled by an abrupt decrease in the energy $\langle e \rangle$, the sharpness of which increases with system size.
	}
\end{figure}

As shown in Fig.~\ref{fig:thermo_111_inset}(b), the order parameter $m_a$ presents a jump which becomes increasingly sharp for larger system sizes.
This trend towards discontinuous behavior (rather than power-law behavior) is suggestive of a first-order phase transition. 
The average energy per spin $\langle e \rangle$, which can be seen in the inset of Fig.~\ref{fig:energy_111_inset}, also displays an abrupt decrease at the transition temperature, the sharpness of which increases for larger systems~\footnote{Note that the difference in ground-state energies per spin for different system sizes is attributable to the different shapes of the boundary at infinity in the lattice summation of the dipolar energies.}. This is consistent with the latent heat expected to accompany a first-order transition.

The associated energy histogram $p(e)$ is shown in Fig.~\ref{fig:energy_111_inset}, at temperatures just above (\mbox{$T/D=0.060$}), approximately at (\mbox{$T/D=0.052$}), and just below (\mbox{$T/D=0.0505$}) the transition (for $L=5$).
Above the transition, there is a single Gaussian-like peak which indicates a unique phase.
The emergence close to the transition temperature of a double-peaked structure indicates the coexistence of two distinct phases of different energies (one of which is the ground state) and thus a first-order transition. 
At lower temperatures, the higher-energy peak disappears as the system increasingly occupies the ground state.

We have examined the scaling of the maximum of the specific heat peak $C_\textrm{max}$ with the system size $L$ but do not find convincing evidence for it scaling with the volume of the system ($\propto L^2$) as expected for a first-order transition (not shown). 
Possible deviations could be due to strong finite-size effects for the modest system sizes that we are able to equilibrate reliably. 
Similar behavior has been found in studies of first-order transitions in long-range interacting Ising spin systems on the square lattice~\cite{osenda_nonequilibrium_2009}.
%
%

\section{Spin relaxation at low temperatures}
\label{sec:spin_relaxation}

\begin{figure}[tp]
\centering
\includegraphics[width=1.0\linewidth]{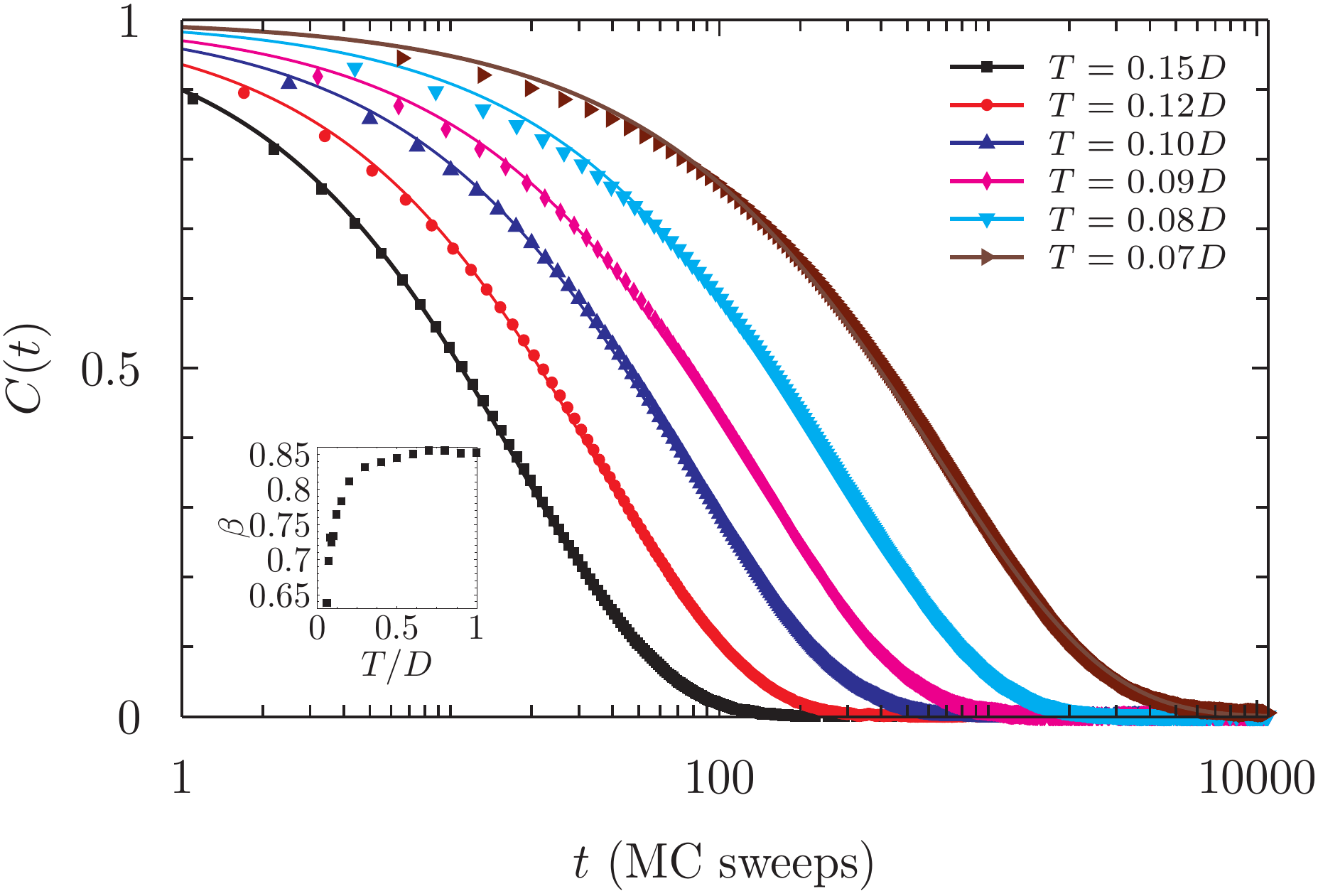}
\\
\vspace{0.5 cm}
\includegraphics[width=1.0\linewidth]{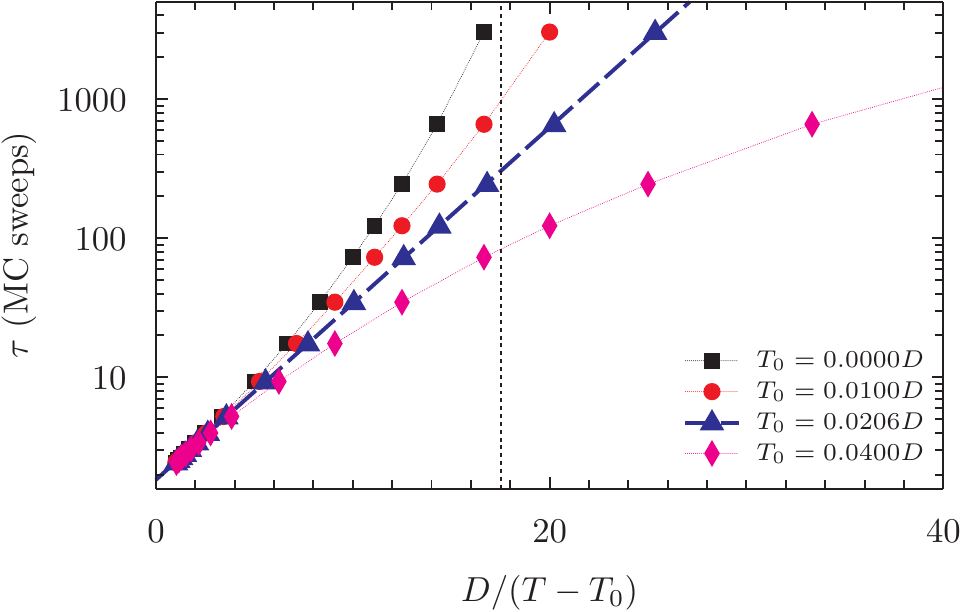}
\caption{\label{fig:L3_autocorr_times_spins} 
Top: 
examples of spin autocorrelation functions (symbols) and relative stretched exponential fits (solid lines), used to extract $\tau$ and $\beta$. The behavior of the latter as a function of temperature is shown in the inset. 
The system size is $L=3$. 
Bottom: 
spin autocorrelation time $\tau$ as a function of inverse temperature (black squares) on a semilogarithmic scale. The noticeable departure from linear scaling at low temperatures is characteristic of super-Arrhenius behavior (black solid squares). The vertical line indicates the finite-size transition temperature $T_c/D \simeq 0.057$. 
The same data are also plotted against $D/(T-T_0)$ to show that $\tau$ diverges according to a Vogel-Fulcher form with \mbox{$T_0/D = 0.0206$} and $\Delta/D = 0.292$ (the dashed blue line is the corresponding fit to the data). 
	}
\end{figure}

Upon approaching the thermodynamic phase transition, the DKIAFM exhibits noticeable freezing 
of its spin dynamics, which we study quantitatively using the spin autocorrelation function, 
\beq
C(t) = \frac{1}{N} \sum_i \sigma_i(0) \sigma_i(t). 
\eeq
We consider single spin-flip dynamics and measure time in Monte Carlo sweeps (that is to say, regular MC sweeps defined as the number of MC spin-flip attempts per spin, in contrast to the modified MC sweeps used in the previous section). We focus on the behavior of the autocorrelation function in thermodynamic equilibrium, equivalent to the \mbox{$t_w\to\infty$} limit of the two-time autocorrelation function $C(t,t_w)$. 
The decay of $C(t)$ is not captured by a simple exponential but is rather described by a stretched exponential, 
\beq
C(t) = \exp[-(t/\tau)^\beta],
\eeq
where $\tau$ is the relaxation timescale and $\beta \leq 1$ is the Kohlrausch exponent. 
Stretched-exponential relaxation is typical of systems with complex energy landscapes and often associated with glassy or supercooled liquid behavior~\cite{cavagna_glass_2003}. 
We fit a stretched exponential to $C(t)$ and extract both the relaxation time $\tau$ and the stretching exponent $\beta$ for different temperatures $T$ (see the top panel of Fig.~\ref{fig:L3_autocorr_times_spins})~\footnote{We note that, for the very lowest temperature included in Fig.~\ref{fig:L3_autocorr_times_spins} (bottom), the proximity to the finite-size transition causes the autocorrelation function to saturate $C(t) \to m^2 > 0$ for $t \gg 1$. Therefore, we fitted instead the connected autocorrelation function $C(t) - m^2$.}. 

The relaxation time $\tau$ for an $L=3$ system obtained from the fit to $C(t)$ is plotted as as a function of inverse temperature $T$ (in units of $D$) in the bottom panel of Fig.~\ref{fig:L3_autocorr_times_spins} (black solid squares). 
The approximate finite-size transition temperature for the $L=3$ system \mbox{$T_c/D = 0.057 \pm 0.002$} is indicated by the vertical dashed line. 
There is clear evidence of super-Arrhenius behavior as the temperature is lowered (above $T_c$). 
The stretching exponent $\beta$ as a function of temperature is shown in the inset of Fig.~\ref{fig:L3_autocorr_times_spins} top panel, demonstrating that the decay of $C(t)$ becomes increasingly stretched at low temperatures. 

Thanks to the two-dimensional nature of the system, we were able to push the numerical simulations to explore a reasonably large range of relaxation timescales. We attempted to fit the temperature dependence using several known forms, and found that only two of them produce good agreement~\footnote{We also considered other forms for the temperature dependence of the relaxation time scale, e.g., critical slowing down $\tau \sim | \frac{D}{T-T_c} |^\alpha$, or \textit{ad hoc}, $\tau \sim \exp[( \Delta / T )^{\alpha} ]$. The corresponding best fits were far less satisfactory, and we do not show them here.}---the Vogel-Fulcher form
\beq
\tau \sim \exp\left(\frac{\Delta}{T-T_0}\right), 
\label{eq:vogel-fulcher}
\eeq
and a parabolic law~\cite{elmatad2009corresponding},
\beq
\tau \sim \exp\left(\frac{A}{T} + \frac{B}{T^2}\right)
\label{eq:inv-T2}
\eeq
(note that both have the same number of fitting parameters). 
Our data show that the former yields a quantitatively better fit, and we focus on it presently. However, the difference is marginal, and for completeness we report a detailed comparison between the two forms in Appendix~\ref{app:VFT_parabolic}. 
Both forms for the relaxation time $\tau$ are characteristic of fragile glass behavior~\cite{cavagna_glass_2003,biroli2013perspective}, which goes hand in hand with the propensity of the system to exhibit supercooled liquid behavior across the thermodynamic transition. 

In the bottom panel of Fig.~\ref{fig:L3_autocorr_times_spins} we plot the timescale $\tau$ on a logarithmic scale as a function of $D/(T-T_0)$ for different values of $T_0$. Our best-fitting parameters are \mbox{$T_0/D = 0.0206 \pm 0.0002$} and \mbox{$\Delta/D = 0.292 \pm 0.001$}. 
Note the long relaxation times ($\tau \sim 10^5$~MC sweeps) at the lowest temperatures.
%
%

\section{Behavior out of equilibrium}
\label{sec:out_of_eq}

\begin{figure}[t!]
\centering	
\includegraphics[width=0.95\columnwidth]{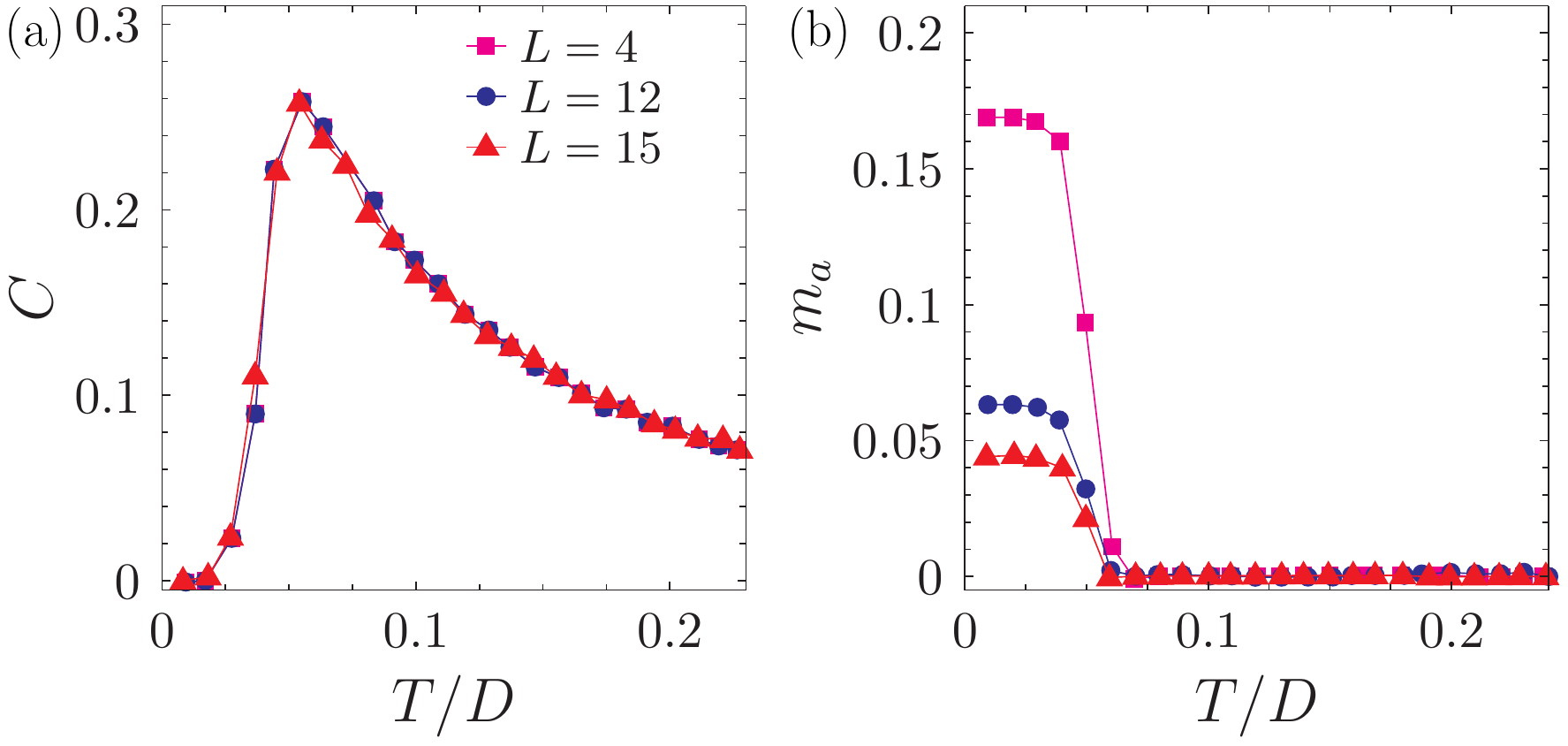}
\caption{\label{fig:out_of_eq} 
Behavior of (a) the specific heat $C$ and (b) the sublattice magnetization $m_a$ for system sizes $L=\{ 4,12,15 \}$ as a function of temperature $T$ in the case where the number of sweeps per temperature step is insufficient to equilibrate the system.
}
\end{figure}

We also ran simulations with larger systems of size $L \in \{ 9, 12, 15 \}$, the features of which we discuss briefly here in regard to out-of-equilibrium behavior. For these system sizes, the cooling protocol described in Sec.~\ref{sec:ground_state} is not sufficient to thermalize the system.
However, the slower the cooling protocol, the more pronounced the peak in the specific heat signaling the incipient transition becomes. To illustrate the development of order out of equilibrium, we simulate systems of sizes $L = \{ 4, 12, 15 \}$ with a protocol that results in substantial (but not complete) ground-state order in the $L=4$ system; we then increase the system size. The specific heat $C$ and order parameter $m_a$ that we identified in this paper (see Sec.~\ref{sec:ground_state})  are illustrated in Figs.~\ref{fig:out_of_eq}(a) and \ref{fig:out_of_eq}(b) respectively. The developing order in the system is most visible in the behavior of the order parameter $m_a$. Even though the value of the order parameter remains rather smaller than the saturated value at all temperatures, it becomes distinctly nonzero---well above statistical fluctuations---at a well-defined temperature that we identify as a reasonable proxy for the thermodynamic transition temperature $T_c$ of the system 
[see Fig.~\ref{fig:out_of_eq}(b)]. The specific heat $C$ behaves in a largely $L$-independent manner and lacks the pronounced $L$-dependent peak present when the system is able to reach equilibrium [see Fig.~\ref{fig:thermo_111_inset}(a)]. This is a signature of the supercooled liquid behavior.

\begin{figure*}[ht]
\centering
\includegraphics[width=0.9\linewidth]{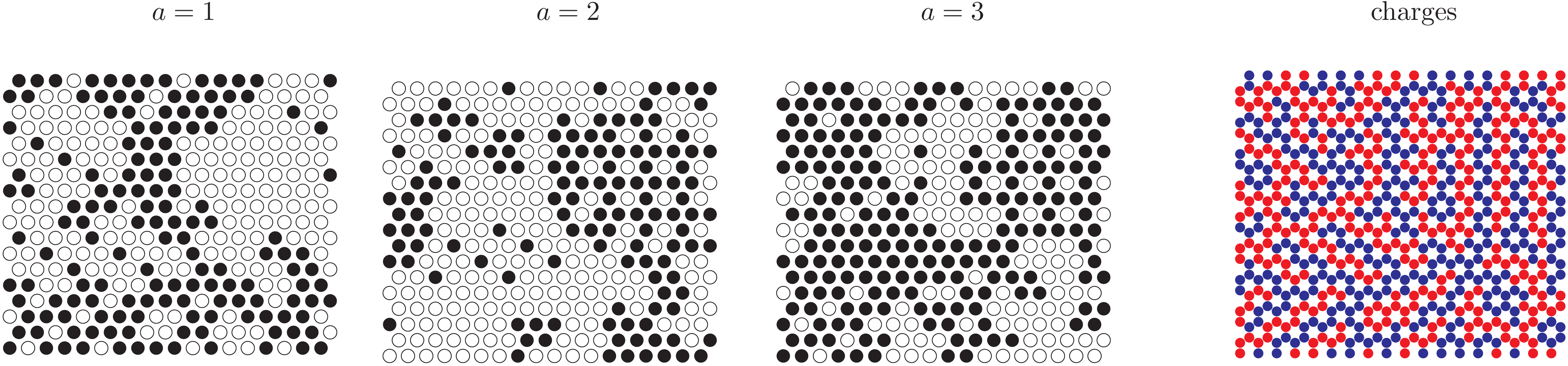}
\caption{\label{fig:L9_spins_sublat}
Individual sublattices $a = \{ 1, 2, 3 \} $ of a low-temperature spin configuration in a $L=9$ system are plotted separately in the first three panels from the left. 
Each sublattice constitutes a triangular lattice.
Regions of full polarization in one sublattice can clearly be seen (for example, the white top-right region for $a=1$) with the corresponding regions in the other sublattices polarized mostly in the opposite direction (mostly black). Note the typical pattern of the partially polarized sublattices with rows of spins that alternate between fully polarized and alternating signs. 
The values for the overall sublattice magnetizations of this configuration are $m_1= -0.074$, $m_2=-0.185$, and $m_3=0.259$. 
The corresponding charge configuration is shown in the right panel, exhibiting a characteristic dendritic stripe pattern. 
	}
\end{figure*}

Understanding the onset of the ordered phase below $T_c$ when the system is out of equilibrium is hindered by the elaborate spin pattern of the 12-site magnetic unit cell. 
We can gain some visual intuition by taking advantage of the fact that one of the sublattices is fully polarized in the ordered state. 
In Fig.~\ref{fig:L9_spins_sublat}, we plot separately the individual sublattices $a=\{ 1, 2, 3\}$ of a low-temperature spin configuration in a $L=9$ system. 
Each sublattice forms a triangular lattice. 
A system which is fully ordered in the ground state [see Fig.~\ref{fig:7_shape_state}(a)] would have one sublattice fully polarized (say, all black) and the two other sublattices partially polarized in the opposite direction (say, mostly white)---with a pattern where one row is fully polarized and the next row has alternating signs. 
This behavior can indeed be recognized in some regions of the system in Fig.~\ref{fig:L9_spins_sublat} (for example, the white top-right region for $a=1$ and corresponding regions for the other sublattices).
By examining individual sublattices in this way, it is clear that the system exhibits some domains consistent with the ground-state order, although identifying boundaries between domains is difficult. 

In the right panel of Fig.~\ref{fig:L9_spins_sublat}, we plot the corresponding configuration of the charges $Q$. 
In the charge picture, it is not immediately easy to identify ordered domains, but on more detailed inspection one can recognize patches of parallel charge stripes, reminiscent of the charge order of the ground state [see Fig.~\ref{fig:7_shape_state}(b)]. 
The different domains with charge stripes oriented along different lattice directions compete with one another, leading to a dendritic arrangement of charge stripes. 
The charge configuration corresponds thus to a kind of ``stripe liquid,'' reminiscent of that observed in a study by Mahmoudian and coworkers~\cite{mahmoudian_glassy_2015} of frustrated Coulomb liquids on the triangular lattice at half-filling (cf., for example, Fig.~2(c) in Ref.~\onlinecite{mahmoudian_glassy_2015}). 
In that work, the authors also observe glassy slow dynamics due to a large manifold of low-lying metastable states; however the divergence of the relaxation timescales at low temperatures in that system is of the more common Arrhenius behavior, characteristic of strong rather than fragile glasses. 
In our system, patches with packed parallel stripes of charges require coordinated ``topological'' (system-spanning) rearrangements of the spin orientations in order to move between low-energy states. It is tantalizing to speculate that an appropriate effective modeling of such spin rearrangements may be key to understanding the glassy slow dynamics (see Appendix~\ref{app:charge_stripe_landscape}). 
%
%

\section{Discussion and conclusions}
\label{sec:conclusion}

To summarize, we have investigated the nature of the ordered phase and phase transition in the DKIAFM. By means of extensive Monte Carlo simulations and the identification of a suitable order parameter, we were able to confirm the ground state proposed in Ref.~\onlinecite{chioar_ground-state_2016}. We also provided evidence that the nature of the transition is first order. 

Interestingly, we notice that a Coulombic system of charges hopping on a kagome lattice~\cite{terao_hopping_2016} appears to exhibit a remarkably similar ordering tendency to the present system, which is also prevented by slow dynamics. We wonder whether the hitherto puzzling ordered state underlying Fig.~1 in Ref.~\onlinecite{terao_hopping_2016} may be the same ordered state demonstrated in our paper. 
Indeed, it may be possible to establish an intuitive connection between the orders exhibited by the two models via the charge mapping discussed in Appendix~\ref{app:charge_stripe_landscape}. 

Upon approaching the phase transition, the DKIAFM exhibits a remarkable propensity to fall out of equilibrium and enter a supercooled liquid phase, avoiding any sign of the full transition altogether~\cite{chioar_ground-state_2016}. We studied the equilibrium behavior of the spin autocorrelation function above the transition and observed that it is well described by a stretched exponential form, typical of glass-forming systems. From the stretched exponential relaxation we obtained the temperature dependence of the equilibrium relaxation time scale and found it to obey a Vogel-Fulcher law, typical of fragile glasses. 

This is a remarkable result in a system without disorder with an eminently simple two-body Hamiltonian in the absence of dynamical constraints (single spin-flip updates). The behavior cannot be related---to the best of our understanding---to the avoided criticality paradigm: the short range interaction terms in the Hamiltonian are frustrated and do not lead per se to a continuous phase transition; moreover, dipolar interactions are not sufficiently long ranged to suppress an ordering transition irrespective of their strength. Interestingly, recent experimental work has hinted that a state similar to a supercooled liquid might exist at low temperatures in the frustrated pyrochlore material \DTO~\cite{kassner2015supercooled}.

Our paper propels the DKIAFM in the study of glassy dynamics in systems without disorder. Further work is needed to understand the origin of the dynamical slowing down---here we merely speculate that it may be related to topological spin rearrangements between low-lying energy states via an effective dumbbell and charge description (discussed in Appendix~\ref{app:charge_stripe_landscape}). What makes this system even more interesting is the potential for experimental verification in several realistic setups from colloidal crystals to artificial nanomagnetic arrays to (layered) bulk kagome materials. 



%
%

\section{Acknowledgements}

We are grateful to J.~P.~Garrahan for insightful discussions on the form of 
the growing relaxation time scale. 
This work was supported, in part, by the  
Engineering and Physical Sciences Research Council Grants No. EP/K028960/1 
and No. EP/M007065/1 (J.H. and C.C.) and by the Deutsche 
Forschungsgemeinschaft via Grant No.~SFB 1143 (R.M.). The calculations were performed using the Darwin Supercomputer of the University of Cambridge High Performance Computing Service.
Statement of compliance with the EPSRC policy framework on
research data: this publication reports theoretical work that does
not require supporting research data.
%
%
\appendix

\section{Quantifying frustration}
\label{app:quantifying_frustration}

In the following we study the level of frustration present in the DKIAFM by considering Pauling estimates of the ground-state degeneracy as the range of the interactions is progressively increased. This illustrates how ordering in the model is expected to arise only from (some) third-neighbor or longer-ranged terms. We also compute the Fourier transform of the full interaction matrix, showing a lowest band that is substantially flatter than the full spectrum bandwidth: another hallmark of a highly frustrated system. 
%
%

\subsection{Pauling estimates for truncated interactions}

\begin{figure}
\centering
\includegraphics[width=0.95\linewidth]{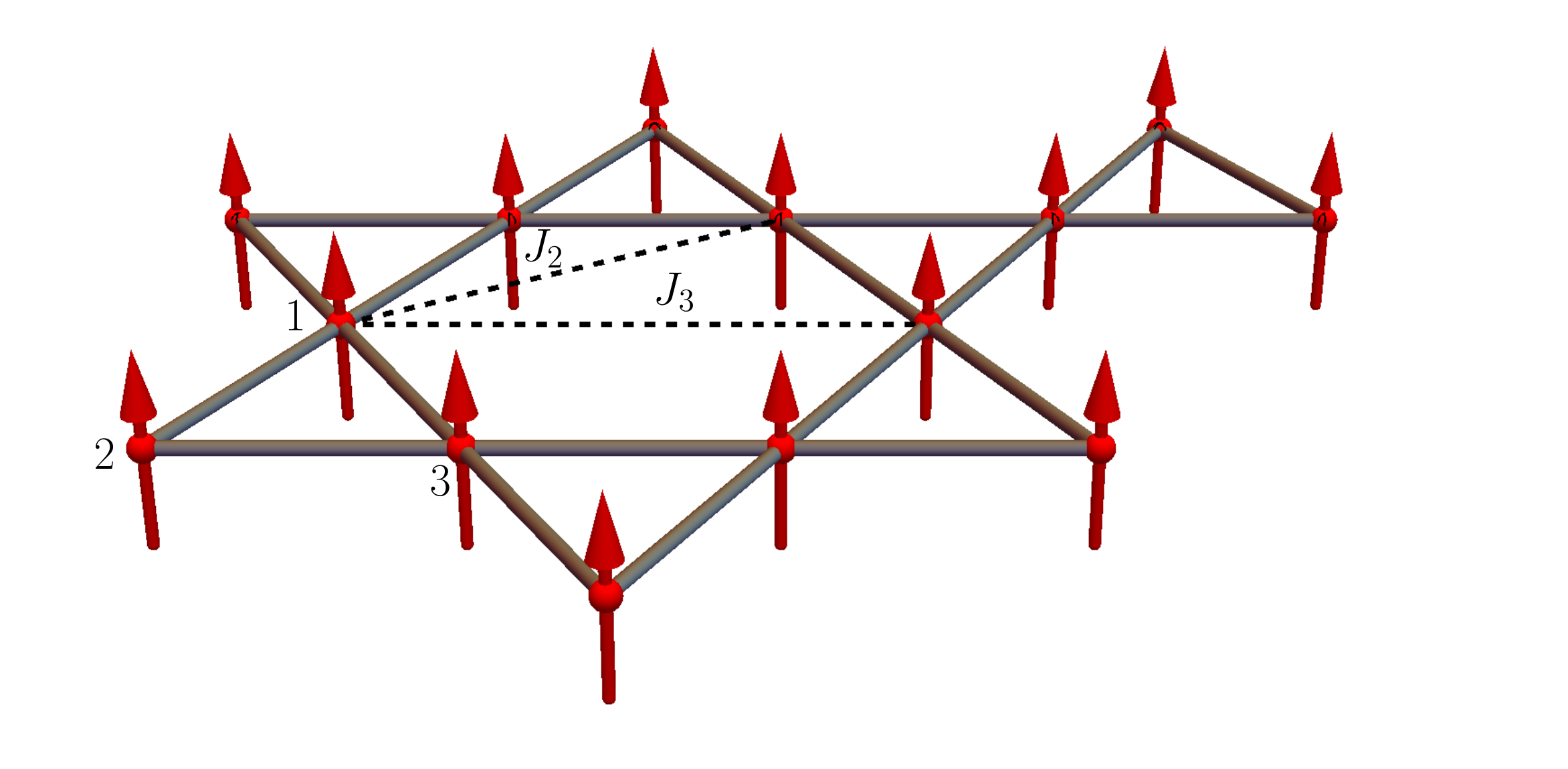}
\caption{Illustration of second- ($J_2$) and third- ($J_3$) neighbor interaction distances on the kagome lattice, indicated by the dashed lines.} 
\label{fig:kagome_lattice_111 pauling}
\end{figure}

We consider here the Hamiltonian~\eqref{eq:dkiafm_hamiltonian} truncated at third-neighbor distance and written for convenience as 
\begin{equation}
	\mathcal{H}  =  
	J_1 \sum_{\langle ij \rangle} \sigma_i \sigma_j 
	+ J_2 \sum_{\langle\langle ij \rangle\rangle} \sigma_i \sigma_j 
	+ J_3 \sum_{\langle\langle\langle ij \rangle\rangle\rangle} \sigma_i \sigma_j.
\label{eq:j1_j1_j3_hamiltonian}
\end{equation}
In accordance with the choice of parameters in Sec.~\ref{sec:model}, we set $J_1=1.5$, $J_2 \simeq 0.692$, and $J_3=0.625$. 
The second- and third-neighbor distances on the kagome lattice are illustrated for convenience in Fig.~\ref{fig:kagome_lattice_111 pauling}. Note that there are two types of third-neighbor distances, whose length is exactly twice the kagome lattice constant: one type is across the hexagonal cells, and the other is along two aligned consecutive bonds (not shown). The $J_3$ term in Eq.~\eqref{eq:j1_j1_j3_hamiltonian} encompasses both types. 

A simple Pauling argument allows to estimate the ground-state degeneracy of the \mbox{$J_1$-$J_2$-$J_3$} model in various regimes. 
For $J_2=J_3=0$, the model reduces to the nearest-neighbor kagome spin ice model of Wills, Ballou and Lacroix~\cite{Wills2002}, for which a Pauling estimate gives an entropy of $\ln[2(3/4)^{2/3}] \simeq 0.5014$ per spin (this is very close to the known exact value $0.5018$~\cite{Kano1953}). 

\begin{figure*}[th!]
\centering
\includegraphics[width=0.9\linewidth]{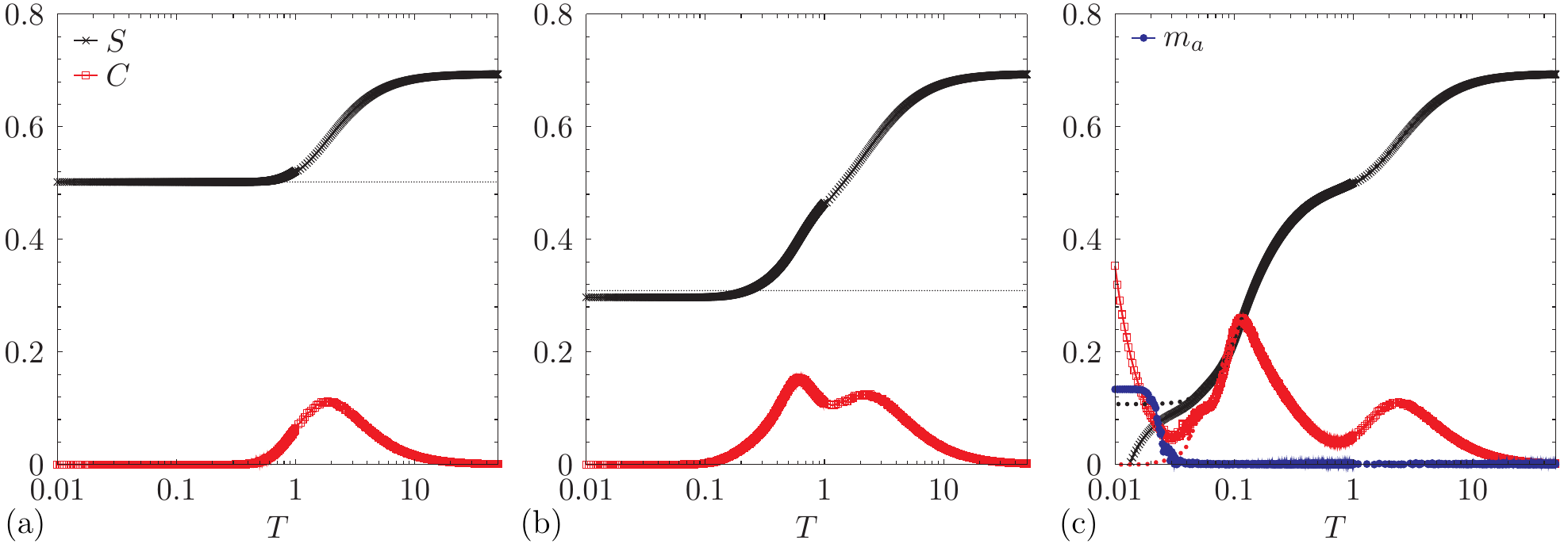}
\caption{
Specific heat $C$ and entropy $S$ of the effective \mbox{$J_1$-$J_2$-$J_3$} model (plotted on the same axis). 
(a) $J_1$ only. 
This model is equivalent to the nearest-neighbor kagome spin ice model.
It exhibits a ground-state entropy in good agreement with the Pauling estimate of $0.501$ per spin (dashed line).
(b) \mbox{$J_1$-$J_2$}. 
This model exhibits a second peak in the specific heat at lower temperatures, associated with the onset of the ice rules on the kagome superlattices dictated by the $J_2$ interactions.
The ground-state entropy is in good agreement with the Pauling estimate of $0.309$ per spin (dashed line).
(c) \mbox{$J_1$-$J_2$-$J_3$}. 
After the second feature in the specific heat, the model falls out of equilibrium, as indicated by the difference between the ensemble-averaged results (points) and time-averaged results (dashed line) for the specific heat and the spin entropy.
Also plotted is the sublattice order parameter $m_a$, which displays an increase from zero, hinting at ordering consistent with the dipolar ground state.
The system size is $L=3$, and the coupling constants where not vanishing are $J_1=1.5$, $J_2 \simeq 0.692$, and $J_3=0.625$. 
}
\label{fig:kagome_single_111_j1_j2_j3_heat_cap_entropy}
\end{figure*}

The $J_2$ interactions form three independent kagome superlattices on which they try to enforce the ice rules (the triangles of these superlattices live inside the hexagons of the original lattice). 
Each kagome superlattice has $N_\text{tri}'=N_\text{tri}/3$ triangles, where $N_\text{tri} = 2 N_s/3$ is the number of triangles in the original kagome lattice ($N_s$ being the total number of spins). 
Therefore the number of possible states can be estimated starting from the nearest-neighbor $J_1$ result as
\begin{eqnarray}
\Omega &\simeq& 
\underbrace{2^{N_s} \times \left( \frac{6}{8} \right)^{N_\text{tri}}}_{\text{kagome ice rule result}} 
\nonumber \\ 
&\times& 
\underbrace{\left( \frac{6}{8} \right)^{N_\text{tri}'} \times \left( \frac{6}{8} \right)^{N_\text{tri}'} \times \left( \frac{6}{8} \right)^{N_\text{tri}'}}_{\text{constraint from three kagome superlattices}} 
\nonumber \\
&=& 2^{N_s} \times \left( \frac{6}{8} \right)^{2N_\text{tri}} 
= 2^{N_s} \times \left( \frac{6}{8} \right)^{4N_s/3} 
\nonumber \\
&\simeq& (1.363)^{N_s} 
\, . 
\end{eqnarray}
This leads to an entropy $\mathcal{S}_{J_1\text{-}J_2} = \ln \Omega \simeq 0.309 N_s$. 

Similarly, the $J_3$ interactions form three triangular superlattices on which they try to enforce the ice rules (these are simply the three sublattices of the original kagome lattice). Each triangular superlattice has $N_s''=N_s/3$ spins and thus $N_\text{tri}''=2 N_s'' = 2 N_s/3 = N_\text{tri}$ triangles. Therefore the number of possible states can be estimated starting with the \mbox{$J_1$-$J_2$} result as
\begin{eqnarray}
\Omega &\simeq& 
\underbrace{2^{N_s} \times \left( \frac{6}{8} \right)^{2 N_\text{tri}}}_{J_1\text{-}J_2  \text{ interactions}} 
\nonumber \\ 
&\times& \underbrace{\left[ \left( \frac{6}{8} \right)^{N_\text{tri}''} \right]^3 }_{\text{constraint from three triangular superlattices}} 
\nonumber \\
&=& 2^{N_s} \times \left( \frac{6}{8} \right)^{10N_s/3} 
\nonumber \\
&\simeq& (0.767)^{N_s} 
\, . 
\label{eq: S J1J2J3}
\end{eqnarray}
This leads to an entropy $\mathcal{S}_{J_1\text{-}J_2\text{-}J_3}	= \ln \Omega \simeq -0.266 N_s$, which is negative and suggests that the system orders. 
(Alternatively, one could use the known residual entropy per spin of a triangular Ising antiferromagnet, \mbox{$\mathcal{S}_{\rm TIAFM} = 0.32306$}, to estimate a Pauling-like reduction factor for each of the three triangular superlattices of $(0.691)^{N_s''}$. Substituting this term inside the square bracket in the second line of Eq.~\eqref{eq: S J1J2J3}, one obtains $\Omega \simeq (0.94)^{N_s}$ and $\mathcal{S}_{J_1\text{-}J_2\text{-}J_3}	\simeq -0.06 N_s$, which is still negative but only very marginally so.) 

It is interesting to notice that, if only a subset of the third-neighbor ($J_3$) interactions are kept (those across the hexagons as illustrated in Figure~\ref{fig:kagome_lattice_111}, but not those along the bonds of the lattice), and if we set \mbox{$J_1=J_2=J_3$}, then the effective model is one where the energy can be written in terms of a sum over all hexagons of the squared magnetization of each hexagon. The ground states of this model have zero total magnetization on each hexagon and a Pauling estimate suggests a residual ground state degeneracy of $\mathcal{S}_{\rm hex}	= (N_s/3) \ln(5/2) \simeq 0.305 N_s$ ($20$ out of the $64$ possible spin arrangements on a hexagon have null magnetization, and there are $N_s/3$ hexagons in a kagome lattice of $N_s$ spins). 
This is an interesting model which might warrant further investigation in the future. 
%
%

\subsection{Simulations}

We investigate the above predictions with Monte Carlo simulations of the $J_1$-$J_2$-$J_3$ Hamiltonian~\eqref{eq:j1_j1_j3_hamiltonian}.

Figure~\ref{fig:kagome_single_111_j1_j2_j3_heat_cap_entropy} shows the results for (a) the specific heat per spin $C$ and entropy $S$ for the $J_1$ only case, (b) the \mbox{$J_1$-$J_2$} case, and (c) the \mbox{$J_1$-$J_2$-$J_3$} case, for a system of size $L=3$. 
As mentioned above, only the $J_1$ case is equivalent to the nearest-neighbor kagome ice model of Wills \emph{et al.}~\cite{Wills2002} and correspondingly displays a broad Schottky peak in the specific heat $C$ at around $T \sim 2$ (in units where $J_1 = 1.5$) signaling the onset of the kagome ice rules and a drop in the entropy $S$ down to a value in good agreement with the Pauling estimate of $0.501$ per spin (dashed line). 

The \mbox{$J_1$-$J_2$} case displays an additional bump in the specific heat at a slightly lower temperature around $T \sim 0.5$ and a drop in the entropy to a value close to the Pauling estimate of $0.309$ per spin (dashed line) 
We therefore ascribe the lower-temperature feature in the specific heat to the onset of the ice rules on the three kagome superlattices.

In the \mbox{$J_1$-$J_2$-$J_3$} case, after the onset of the kagome ice rules, the second feature in the specific heat is pushed to lower temperatures, and the system falls out of equilibrium, signaled by the difference between the ensemble-averaged results and the purely time-averaged results. 
The divergence of the specific heat at low temperatures in the ensemble-averaged case indicates that the system does not find a unique energy minimum despite us using extremely slow annealing protocols at low temperatures as in the study of the full dipolar case above (for concreteness, we cool from $T = 1$ in decrements of  $ 5 \times 10^{-4}$ using \mbox{$2 \times 10^4$} modified MC sweeps for equilibration at each temperature step). 

Despite our inability to equilibrate the system, we do find a small signature of a trend towards the proposed ground state. The order parameter $m_a$ from Eq.~\eqref{eq:sublattice_magnetisation} remains zero up to fluctuations for the \mbox{$J_1$-$J_2$} case at all temperatures, whereas for the \mbox{$J_1$-$J_2$-$J_3$} case it increases from zero to a value of about $0.13$ for temperatures lower than approximately $T = 0.03$, signaling some development of order consistent with the proposed ground state [see Fig.~\ref{fig:kagome_single_111_j1_j2_j3_heat_cap_entropy} (c)]. 
However, other types of order are consistent with this signature, and further work is needed to say anything conclusive on the matter. It seems that the \mbox{$J_1$-$J_2$-$J_3$} model is perhaps even more frustrated than the full dipolar model and that further-neighbor interactions play a key role in relieving some frustration and selecting the ground state. 
%
%

\begin{figure}[t!]
\centering
\includegraphics[width=0.9\linewidth]{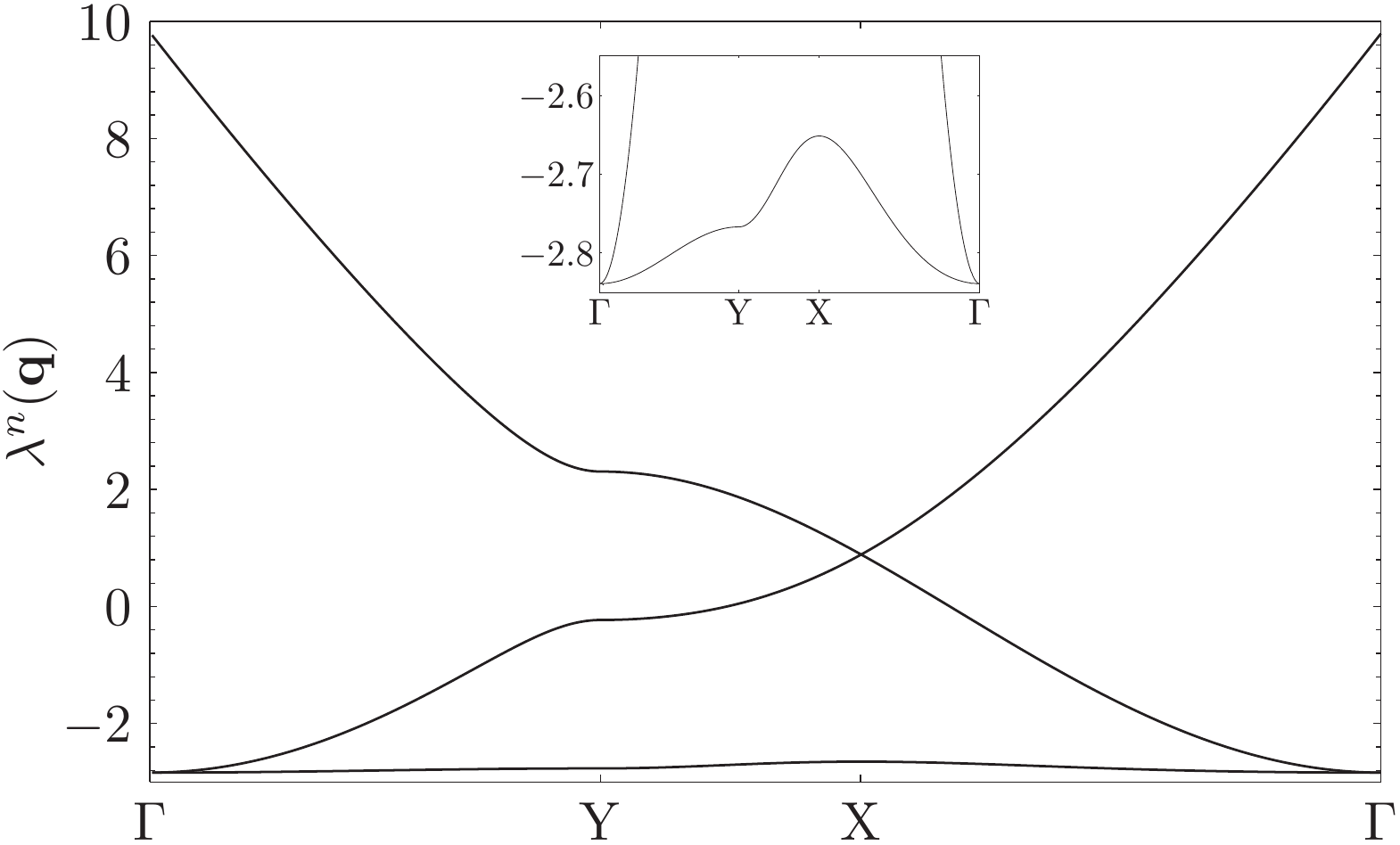}
\caption{\label{fig:ewald_ft}  
Eigenvalues $\lambda^n(\mathbf{q})$ of the Fourier transform of the interaction matrix $J^{ab}(\mathbf{q})$ on a path in the Brillouin zone \mbox{$\Gamma \rightarrow \text{Y} \rightarrow \text{X} \rightarrow \Gamma$}. 
There are three branches due to the three sites in the unit cell. 
The flatness of the bottom branch is characteristic of frustration. 
Inset: zoom on the bottom branch $\lambda^\mathrm{min}(\mathbf{q})$, which presents a minimum at the $\Gamma$ point. 
}
\end{figure}
			
\subsection{Interaction matrix}

Each lattice site $i \equiv (l,a)$ has an index $l$ which labels the sites of the Bravais lattice formed by the centers of the up-type kagome triangles and an index $a \in \{1,2,3\}$ which labels the sublattice (see Fig.~\ref{fig:kagome_lattice_111}). 
Namely, the spins $\mathbf{S}_i \equiv \mathbf{S}_l^a \equiv \mu \sigma_l^a \hat{\mathbf{e}}_z$ have positions $\mathbf{r}_i \equiv \mathbf{r}_l^a \equiv \mathbf{R}_l+\mathbf{e}^a$, where $\{ \mathbf{R}_l \}$ point to the centers of the up triangles, and $\{ \mathbf{e}^a \}$ are the vectors from the centers of the triangles to each of the three spins, 
\begin{eqnarray}
\mathbf{e}^1 &=& (0,1/\sqrt{3},) 
\\ 
\mathbf{e}^2 &=& (-1,-1/\sqrt{3})/2, 
\\ 
\mathbf{e}^3 &=& (1,-1/\sqrt{3})/2, 
\end{eqnarray}
in units of the kagome lattice constant. 

The Hamiltonian~\eqref{eq:dkiafm_hamiltonian} can then be written as 
\begin{equation}
\mathcal{H} = \sum_{l m } \sum_{a b} 
  J^{a b}(\mathbf{R}_{lm}) \sigma_l^a \sigma_m^b 
\, , 
\end{equation}
where $\mathbf{R}_{lm} \equiv \mathbf{R}_m - \mathbf{R}_l$. In Fourier space,
\begin{equation}
	\mathcal{H} = \sum_\mathbf{q} \sum_{a b} J^{a b}(\mathbf{q}) \sigma_\mathbf{q}^a \sigma_\mathbf{-q}^b,
\end{equation}
where 
$
\sigma^a_l = \sum_\mathbf{q} \sigma^a_\mathbf{q} 
  \exp(i \mathbf{q} \cdot \mathbf{r}_l^a)$ 
and 
\begin{equation}
J^{a b}(\mathbf{R}_{lm}) = \sum_\mathbf{q} 
  J^{a b}(\mathbf{q}) \exp(i \mathbf{q} \cdot \mathbf{r}_{lm}^{ab})
\end{equation}
is the $3 \times 3$ interaction matrix, where $\mathbf{r}_{lm}^{ab} \equiv \mathbf{r}_m^b - \mathbf{r}_l^a$.

The eigenvalue spectrum $\lambda^n(\mathbf{q})$ of 
$J^{a b}(\mathbf{q})$ is shown in Fig.~\ref{fig:ewald_ft}. 
It has three branches due to the three sites in the unit cell.
The flatness of the bottom branch $\lambda^\text{min}(\mathbf{q})$ (illustrated in detail in the inset of Fig.~\ref{fig:ewald_ft}) is characteristic of frustration in the model. Note that its bandwidth is only about 2\% of that of the full spectrum. 
The minimum at the $\Gamma$ point suggests that, at the mean-field level, the leading ordering instability from the high-temperature phase is expected to be at $\mathbf{q}^* = (0,0)$. 
(Our results are the extension to the Ising case of the results found for Heisenberg spins by Maksymenko and co-workers~\cite{maksymenko_classical_2015}). 
%
%

\section{Vogel-Fulcher vs parabolic fit}
\label{app:VFT_parabolic}

\begin{figure}[t!]
\centering
\includegraphics[width=0.9\linewidth]{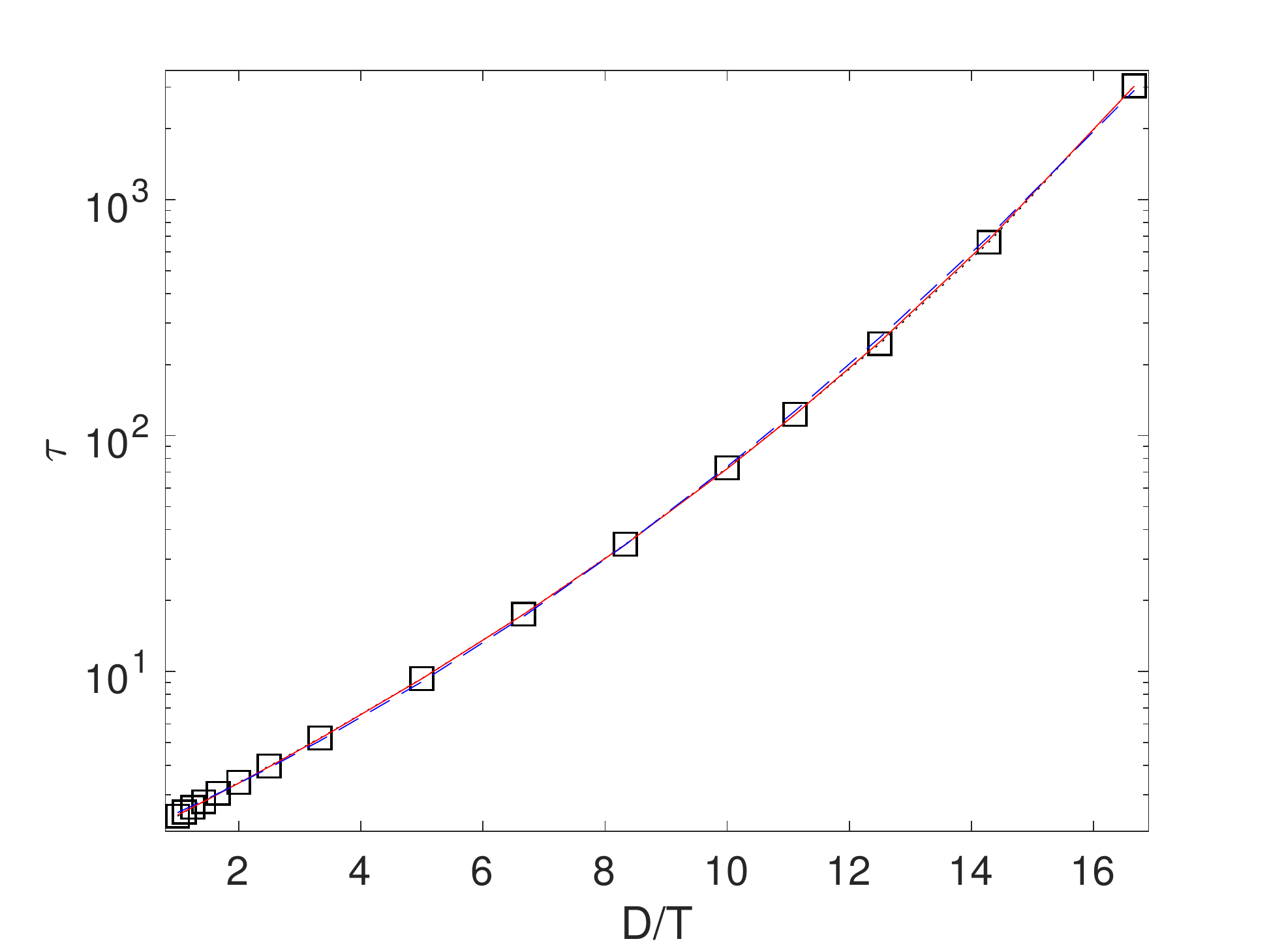}
\caption{\label{fig:VFT_parabolic_fit}  
Best fits to the dependence of the relaxation time $\tau$ (squares) vs inverse temperature $T$ using Vogel-Fulcher (red solid line) and parabolic (blue dashed line) forms. The Vogel-Fulcher form provides a better fit to the data.
}
\end{figure}

In order to compare the temperature dependence of the relaxation time scale in the DKIAFM to the Vogel-Fulcher and parabolic laws, we fit $\ln\tau(T)$ to 
\begin{eqnarray}
\ln\tau_0 + \frac{\Delta}{T - T_0},
\end{eqnarray}
and to 
\begin{eqnarray}
\ln\tau_0 + \frac{A}{T} + \frac{B}{T^2}, 
\end{eqnarray}
with three fitting parameters each. The best fits give 
\mbox{$\tau_0 \simeq 1.83$}, \mbox{$\Delta/D \simeq 0.292$}, \mbox{$T_0/D \simeq 0.0206$}, 
and 
\mbox{$\tau_0 \simeq 1.94$}, \mbox{$A/D \simeq 0.252$}, \mbox{$B/D^2 \simeq 0.0112$}, 
respectively, and are shown in Fig.~\ref{fig:VFT_parabolic_fit}. 
%
%

%
%
We use the full range of numerical values of $\tau(T)$ for the fit, and we find a marginally better result using the Vogel-Fulcher form. This can be quantified by computing the squared difference between the best fit and the numerical data, summed over all temperature data points; the resulting variance is $3.5$ times larger for the parabolic law than for the Vogel-Fulcher form. The difference is however visibly marginal as demonstrated by the comparison in Fig.~\ref{fig:VFT_parabolic_fit}. 
%
%

\section{Effective charge picture, emergent charge stripes, and freezing}
\label{app:charge_stripe_landscape}

To better understand the nature of the low-energy states in the DKIAFM, it is interesting to draw a parallel with a related model: kagome ice~\cite{Wills2002}. 
In the latter, the Ising spins lie within the plane of the lattice, and point directly into or out of a triangle. 
A useful way for understanding kagome ice derives from the so-called dumbbell picture where each spin is represented as a pair of magnetic charges $\pm q$ separated by a distance $a$ such that $\mu = qa$~\cite{castelnovo_magnetic_2008,Chern2011}. 
Specifically, it is customary to choose $a$ so that the three charges in each triangle of the kagome lattice meet precisely at its center. 
\begin{figure}[tp!]
\centering
\includegraphics[width=1.0\linewidth]{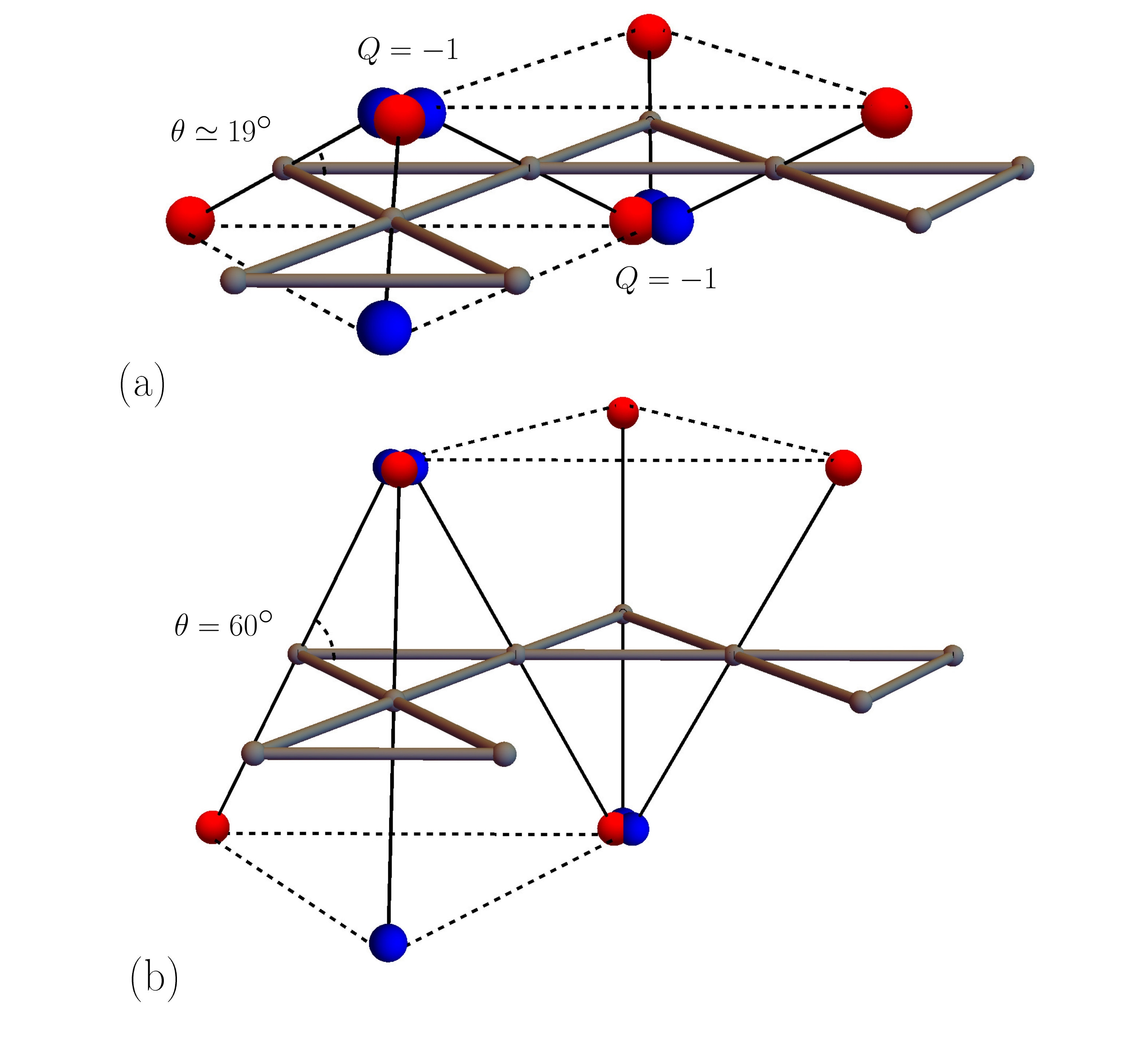}
\caption{\label{fig:dumbbells} 
Each spin of magnitude $\mu$ is decomposed into a dumbbell, i.e., a pair of charges of strength $\pm q$ separated by a distance $a$ such that $\mu=qa$. 
Starting in-plane charges, with the distance $a$ chosen so that the charges at the end of the dumbbells overlap at the centers of the triangles, we progressively tilt the spins out of the plane of the lattice. In the process, we increase $a$ and correspondingly reduce $q$ so that the charges remain overlapping and $\mu=qa$ is kept constant. 
The sum of three overlapping charges is proportional to the corresponding triangular charge $Q_\alpha$ introduced in the main text (and shown as $Q$ in the figure). 
The limiting case of spins perpendicular to the kagome plane corresponds to the DKIAFM, which can then be seen as two infinitely separated triangular layers of charges (indicated by the dashed lines). 
	}
\end{figure}
To leading order, the Hamiltonian of kagome ice can then be written in terms of the total charges $Q_\alpha$ inside each triangle (labeled by $\alpha$), 
\begin{equation}
\mathcal{H}_\text{eff} = 
\sum_\alpha 
  \frac{1}{2} v_0 Q_\alpha^2 
	+ 
	\frac{\mu_0}{8 \pi} \sum_{\beta < \alpha} 
	\frac{Q_\alpha Q_\beta}{r_{\alpha \beta}} 
\, . 
\label{eq:dumbbell_hamiltonian}
\end{equation}
The first term is a chemical potential of strength $v_0$ for the charges, and the second term is a long-range Coulomb interaction between them. 
Notice that the lattice formed by the centres of the triangles is a honeycomb lattice dual to the original kagome lattice. 
Much of the physics of kagome ice can be understood more intuitively in terms of such system of interacting charges than in terms of the original spins. 

\begin{figure}[t!]
\centering
\includegraphics[height=0.5\columnwidth]{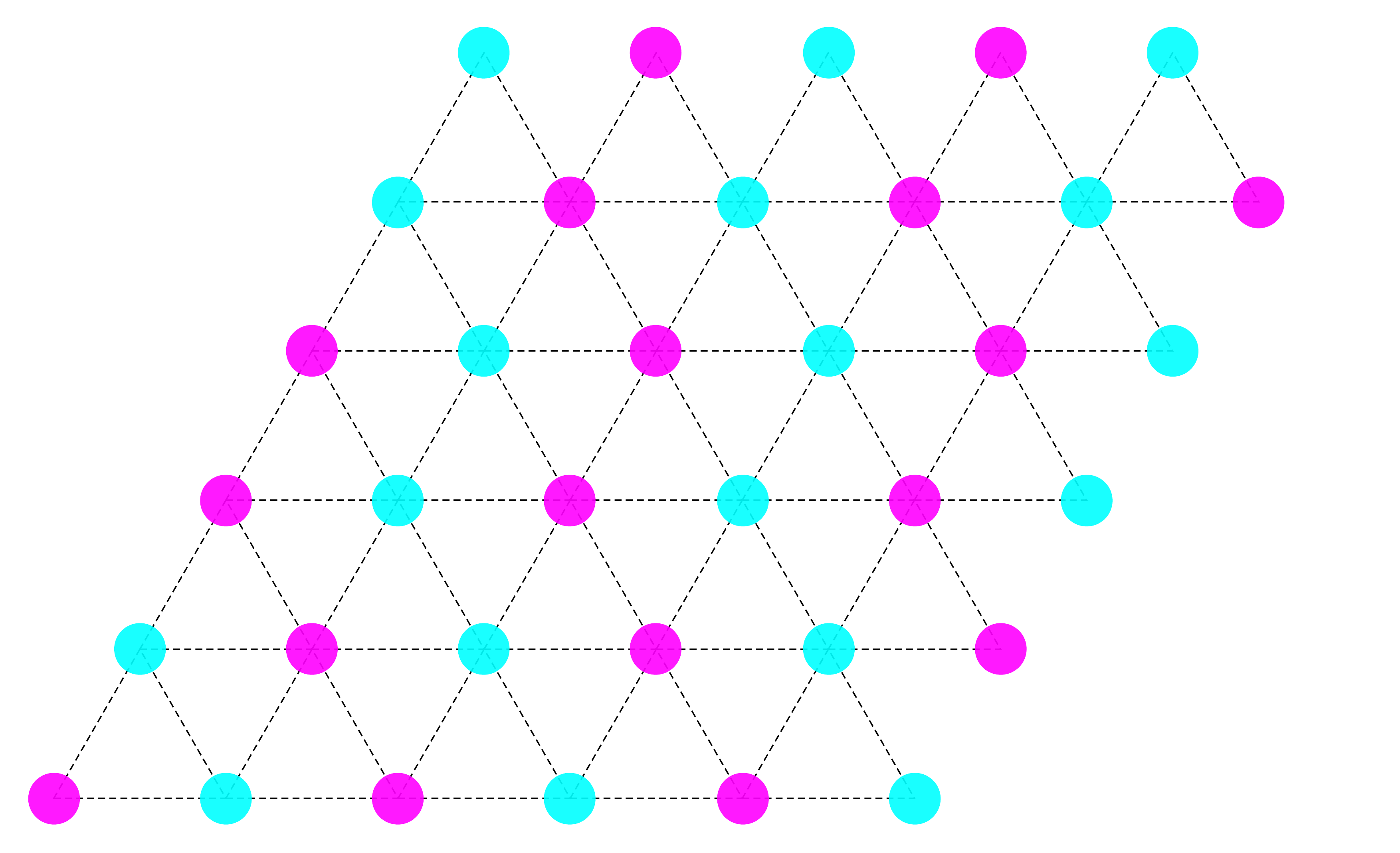}
\\
\vspace{0.4 cm}
\includegraphics[height=0.5\columnwidth]{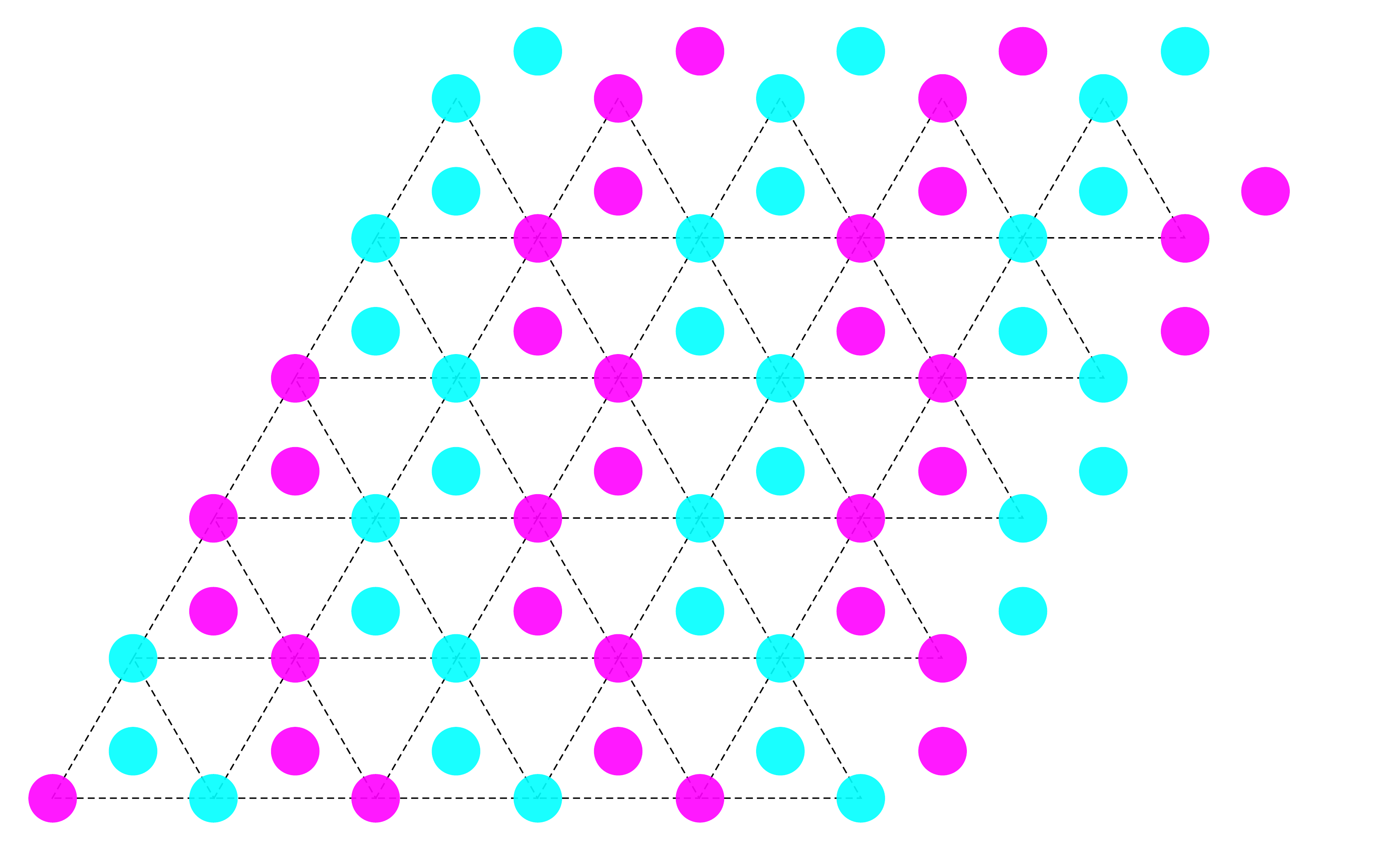}
\caption{\label{fig:triangular_lattice_stripes}
Coulomb-interacting positive (magenta) and negative (cyan) charges on the triangular lattice are expected to order in patterns, such as that in the top panel, with charges alternating along one lattice direction and random along the other. Note that one can view the charge pattern in terms of stripes along the lattice bonds (running largely from top to bottom in the configuration chosen here). 
The interactions between the two layers, dictated by the dipolar coupling between the underlying spins, favors like charges close to one another across layers, and leads to an overall charge arrangement, such as the one in the bottom panel, also exhibiting charge stripes. 
There is a whole class of these energetically low-lying states, which includes the dipolar ground state. 
}
\end{figure}

Inspired by the dumbbell construction of the charges in kagome ice, one can view the DKIAFM as the limit where the spins are progressively tilted until they become perpendicular to the kagome plane (as illustrated pictorially in Fig.~\ref{fig:dumbbells}). 
In the process, one ought to take the limit $a \rightarrow \infty$ (and, correspondingly, $q \rightarrow 0$) to preserve the charges at the end of the dumbbells overlapping at the same location. 
(Note that the magnetic charges $Q_\alpha$ associated with each triangle in the dumbbell picture are, in fact, proportional to the charges 
$Q_{\bigtriangleup}$ 
and 
$Q_{\bigtriangledown}$ 
introduced in Sec.~\ref{sec:ground_state}.) 
Of course, the greater the tilt, the less accurate the dumbbell picture becomes, and at some point the description in terms of resummed charges ought to break down. However, we are tempted to ignore this issue and see what the naive limiting scenario suggests about the behavior of the system. 
Resumming the charges as above leads to a description in terms of two triangular layers (one formed by the centers of the up-kagome triangles; the other by the centers of the down-kagome triangles, shown as dashed lines in Fig.~\ref{fig:dumbbells}) that, to first approximation, are decoupled from one another. 
The coupling within each triangular layer is due to the (antiferromagnetic) Coulomb interaction between charges given by the second term in the effective Hamiltonian~\eqref{eq:dumbbell_hamiltonian}. 
Such a Coulomb-interacting system on the triangular lattice is predicted~\cite{lee_dense_1991, lee_phase_1992} to be partially frustrated and have ground states where charges alternate in one lattice direction but are random in the other direction. 
An example is illustrated in Fig.~\ref{fig:triangular_lattice_stripes} (top panel). 
These states can be viewed as charge-stripe patterns on the triangular lattice with alternating lines of like charges that correspond to the path of a random walk that can either turn left or right as it moves vertically from one row to the next. 
There is a whole family of $\sim 2^L$ such states, each corresponding to a particular choice of path for the stripes. 

The two triangular layers appear decoupled in terms of the Coulomb interaction between the resummed charges $Q_\alpha$, which, in the limit of dumbbells perpendicular to the plane, are infinitely separated from one another. However, the dipolar interaction between the original spins can be seen to favor like charges to sit close to one another across layers. 
Indeed, a charge in one layer adjacent to a charge of the same sign in the other layer corresponds to a mainly antiferromagnetic (and thus energetically favored) spin arrangement as illustrated in Fig.~\ref{fig:dumbbells}(b). 
If we pair the top and bottom triangular charge layers, each in one of their stripe configurations, so as to maximize the proximity between like charges across layers, we obtain overall charge arrangements, such as the one illustrated in Fig.~\ref{fig:triangular_lattice_stripes} (bottom panel). 
One triangular layer becomes a slave to the other, but an $\sim 2^L$ degeneracy survives and it again takes the form of charge stripes randomly turning left and right as they stretch across the lattice. 
The total number of states in this family is thus subextensive, i.e., its entropy scales with the linear size of the system. 
Remarkably, one of these states is indeed the {\sf 7}-shape ground state of the DKIAFM, illustrated in Fig.~\ref{fig:7_shape_state}(b). 

Although we have clearly taken the dumbbell picture and corresponding charge representation well beyond its limit of validity and the energetic arguments above cannot be trusted per se, one can compute the actual energies of various charge-stripe configurations, such as that in Fig.~\ref{fig:triangular_lattice_stripes} (bottom panel) in terms of original spins via the Hamiltonian~\eqref{eq:dkiafm_hamiltonian}. 
We find that many of them lie very close in energy to the ground state (with an energy difference of as little as 1.3\%), whilst differing from it in configuration space by a topological rearrangement of at least $O(L)$ spins. 
Indeed, in order to change a charge stripe state into another without introducing costly defects (dislocations, namely, stripe endpoints and branching), one needs to modify the spin configuration so as to move the charge stripes consistently across the whole system, which amounts to 
a system-spanning topological update. 

We speculate that the existence of this subextensive manifold of energetically low-lying but configurationally topologically distinct states may be one of the key reasons underpinning the strong freezing observed in the DKIAFM at low temperatures. 
We stress that this is a mere speculation, and, in particular, we make no claim 
to have identified all low-lying energy states, which may well be extensive 
in number, as more typically expected in glassy systems. 
%
%

%

\end{document}